\documentclass[twocolumn,showpacs,preprintnumbers,amsmath,amssymb,superscriptaddress]{revtex4}

\usepackage{graphicx}
\usepackage{dcolumn}
\usepackage{bm}

\def\bea{\begin{eqnarray}}
\def\eea{\end{eqnarray}}

\begin{document}


\title{Two-particle correlations on transverse momentum and momentum dissipation in Au-Au collisions at $\sqrt{s_{NN}}$ = 130 GeV}

%

\affiliation{}
\affiliation{Argonne National Laboratory, Argonne, Illinois 60439}
\affiliation{University of Birmingham, Birmingham, United Kingdom}
\affiliation{Brookhaven National Laboratory, Upton, New York 11973}
\affiliation{California Institute of Technology, Pasadena, California 91125}
\affiliation{University of California, Berkeley, California 94720}
\affiliation{University of California, Davis, California 95616}
\affiliation{University of California, Los Angeles, California 90095}
\affiliation{Carnegie Mellon University, Pittsburgh, Pennsylvania 15213}
\affiliation{Creighton University, Omaha, Nebraska 68178}
\affiliation{Nuclear Physics Institute AS CR, 250 68 \v{R}e\v{z}/Prague, Czech Republic}
\affiliation{Laboratory for High Energy (JINR), Dubna, Russia}
\affiliation{Particle Physics Laboratory (JINR), Dubna, Russia}
\affiliation{University of Frankfurt, Frankfurt, Germany}
\affiliation{Institute of Physics, Bhubaneswar 751005, India}
\affiliation{Indian Institute of Technology, Mumbai, India}
\affiliation{Indiana University, Bloomington, Indiana 47408}
\affiliation{Institut de Recherches Subatomiques, Strasbourg, France}
\affiliation{University of Jammu, Jammu 180001, India}
\affiliation{Kent State University, Kent, Ohio 44242}
\affiliation{Institute of Modern Physics, Lanzhou, China}
\affiliation{Lawrence Berkeley National Laboratory, Berkeley, California 94720}
\affiliation{Massachusetts Institute of Technology, Cambridge, MA 02139-4307}
\affiliation{Max-Planck-Institut f\"ur Physik, Munich, Germany}
\affiliation{Michigan State University, East Lansing, Michigan 48824}
\affiliation{Moscow Engineering Physics Institute, Moscow Russia}
\affiliation{City College of New York, New York City, New York 10031}
\affiliation{NIKHEF and Utrecht University, Amsterdam, The Netherlands}
\affiliation{Ohio State University, Columbus, Ohio 43210}
\affiliation{Panjab University, Chandigarh 160014, India}
\affiliation{Pennsylvania State University, University Park, Pennsylvania 16802}
\affiliation{Institute of High Energy Physics, Protvino, Russia}
\affiliation{Purdue University, West Lafayette, Indiana 47907}
\affiliation{Pusan National University, Pusan, Republic of Korea}
\affiliation{University of Rajasthan, Jaipur 302004, India}
\affiliation{Rice University, Houston, Texas 77251}
\affiliation{Universidade de Sao Paulo, Sao Paulo, Brazil}
\affiliation{University of Science \& Technology of China, Hefei 230026, China}
\affiliation{Shanghai Institute of Applied Physics, Shanghai 201800, China}
\affiliation{SUBATECH, Nantes, France}
\affiliation{Texas A\&M University, College Station, Texas 77843}
\affiliation{University of Texas, Austin, Texas 78712}
\affiliation{Tsinghua University, Beijing 100084, China}
\affiliation{Valparaiso University, Valparaiso, Indiana 46383}
\affiliation{Variable Energy Cyclotron Centre, Kolkata 700064, India}
\affiliation{Warsaw University of Technology, Warsaw, Poland}
\affiliation{University of Washington, Seattle, Washington 98195}
\affiliation{Wayne State University, Detroit, Michigan 48201}
\affiliation{Institute of Particle Physics, CCNU (HZNU), Wuhan 430079, China}
\affiliation{Yale University, New Haven, Connecticut 06520}
\affiliation{University of Zagreb, Zagreb, HR-10002, Croatia}

\author{J.~Adams}\affiliation{University of Birmingham, Birmingham, United Kingdom}
\author{M.M.~Aggarwal}\affiliation{Panjab University, Chandigarh 160014, India}
\author{Z.~Ahammed}\affiliation{Variable Energy Cyclotron Centre, Kolkata 700064, India}
\author{J.~Amonett}\affiliation{Kent State University, Kent, Ohio 44242}
\author{B.D.~Anderson}\affiliation{Kent State University, Kent, Ohio 44242}
\author{M.~Anderson}\affiliation{University of California, Davis, California 95616}
\author{D.~Arkhipkin}\affiliation{Particle Physics Laboratory (JINR), Dubna, Russia}
\author{G.S.~Averichev}\affiliation{Laboratory for High Energy (JINR), Dubna, Russia}
\author{Y.~Bai}\affiliation{NIKHEF and Utrecht University, Amsterdam, The Netherlands}
\author{J.~Balewski}\affiliation{Indiana University, Bloomington, Indiana 47408}
% ...

\collaboration{STAR Collaboration}\noaffiliation


\date{\today}

\begin{abstract}
Measurements of two-particle correlations on transverse 
momentum $p_t$ for Au-Au collisions at $\sqrt{s_{NN}} = 130$~GeV are presented. Significant large-momentum-scale correlations are observed
for charged primary hadrons with $0.15 \leq p_t \leq 2$~GeV/$c$ and
pseudorapidity $|\eta| \leq 1.3$.  Such correlations
were not observed in a similar study at lower energy and are not
predicted by theoretical collision models. Their direct relation to  mean-$p_t$ fluctuations measured in the same angular acceptance is demonstrated. 
Positive correlations are observed for pairs of particles which have large $p_t$
values while negative correlations occur for pairs in which one particle has
large $p_t$ and the other has much lower $p_t$. The correlation amplitudes
per final state particle increase with collision centrality.
The observed correlations are consistent with a scenario in which 
the transverse momentum of hadrons associated with initial-stage
semi-hard parton scattering
is dissipated by the medium to lower $p_t$.
\end{abstract}


\maketitle

\section{Introduction}
\label{Sec:Intro}

Studying two-particle correlations and event-wise fluctuations can provide essential information about the medium produced in ultrarelativistic heavy ion collisions~\cite{stock,poly,tricrit}. At the collision energies available at the Relativistic Heavy Ion Collider (RHIC) energetic parton scattering occurs at sufficient rate to enable quantitative studies of in-medium modification of parton scattering and the distribution of correlated charged hadrons associated with
those energetic partons.  Modification of those correlation structures is expected as the bulk medium produced in ultrarelativistic heavy ion collisions increases in spatial extent and
energy density with increasing collision centrality. Analyses of the centrality dependence in Au-Au collisions of high-$p_t$ back-to-back jet angular correlations based on a leading-particle technique ({\em e.g.,} leading-particle $p_t > 4$ GeV/$c$, associated particle $p_t < 4$ GeV/$c$) reveal strong suppression for central collisions~\cite{backjet,phenixbackjet}, suggesting the development of a medium which dramatically dissipates momentum.  
Complementary studies of the lower-$p_t$ bulk medium, its correlation structure on transverse momentum, and how those correlations evolve with collision centrality provide a measure of the momentum transport from the few GeV/$c$ range to
lower $p_t$ of order a few tenths of a GeV/$c$ where the bulk hadronic
production occurs. 
Such studies are an essential part of understanding the
nature of the medium produced in heavy ion collisions at RHIC.


In addition to jet angular correlations at high-$p_t$
substantial nonstatistical fluctuations in event-wise mean transverse momentum $\langle p_t \rangle$ of charged particles from Au-Au collisions were reported by the STAR~\cite{meanptprl} and PHENIX~\cite{phenixmeanpt} experiments at RHIC. $\langle p_t \rangle$ fluctuations at RHIC are much larger than those reported at the CERN Super Proton Synchrotron (SPS) with one-tenth the CM energy~\cite{gunter}, and were not predicted by theoretical models~\cite{meanptprl,jetquench,rqmd,ampt}. $\langle p_t \rangle$ fluctuations could result from several sources
including collective flow ({\em e.g.,} elliptic flow~\cite{starflow} when azimuthal acceptance is incomplete), local temperature fluctuations, quantum interference~\cite{starhbt},
final-state interactions, resonance decays, longitudinal fragmentation~\cite{lund},
and initial-state multiple scattering~\cite{iss} including hard parton scattering~\cite{jetquench} with subsequent in-medium dissipation~\cite{newref}.
$\langle p_t \rangle$ fluctuations can be directly related to integrals of two-particle correlations over the $p_t$ acceptance.  Correlations on $p_t$, by providing differential information, better reveal the underlying dynamics for the observed nonstatistical fluctuations in $\langle p_t \rangle$.

In this paper we report the first measurements at RHIC of two-particle correlations (based on number of pairs) on two-dimensional (2D) transverse momentum space $(p_{t1},p_{t2})$ for all charged particles with $0.15 \leq p_t \leq 2$ GeV/$c$ and $|\eta| \leq 1.3$ (pseudorapidity) using the $\sqrt{s_{NN}} = 130$~GeV Au-Au collisions observed with the STAR detector~\cite{star}. 
This analysis is intended to reveal the response of the bulk medium to strong momentum dissipation and probe the dynamical origins of $\langle p_t \rangle$ fluctuations. 
The data used in this analysis are described in Sec.~\ref{Sec:Data} and the
analysis method, corrections and errors are discussed in
Sec.~\ref{Sec:Analysis}. Models and fits to the data are presented in
Secs.~\ref{Sec:Dist} and \ref{Sec:Model}, respectively. A discussion and summary
are presented in the last two sections~\ref{Sec:Diss} and \ref{Sec:Sum}.

\section{Data}
\label{Sec:Data}

Data for this analysis were obtained with the STAR detector~\cite{star}
using a 0.25~T uniform magnetic field parallel to the beam axis.
A minimum-bias event sample (123k triggered events) required coincidence of two Zero-Degree Calorimeters (ZDC); a 0-15\% of total cross section event sample (217k triggered events) was defined by a threshold on the Central Trigger Barrel (CTB) scintillators, with ZDC coincidence. Event triggering and charged-particle measurements with the Time Projection Chamber (TPC) are described in \cite{star}.  Approximately 300k events were selected for use in this analysis.
A primary event vertex within 75~cm of the axial center of the TPC was required.
Valid TPC tracks fell within the detector acceptance used here, defined by $0.15 < p_t < 2.0$~GeV/$c$, $|\eta| < 1.3$ and $2 \pi$ in azimuth. Primary tracks were defined as having a distance of closest approach less than 3~cm from the reconstructed primary vertex which included a large fraction of true primary hadrons plus approximately 7\% background contamination~\cite{spectra} from weak decays
and interactions with the detector material. In addition accepted particle tracks were required to include a minimum of 10 fitted points (the TPC contains 45 pad rows in
each sector) and, to eliminate split tracks ({\em i.e.,} one
particle trajectory reconstructed as two or more tracks), the fraction of space points used in a track fit relative to the maximum number expected was required to be $> 52\%$.
Particle identification was not implemented but charge sign was determined.
Further details associated with track definitions, efficiencies and quality cuts are described in~\cite{spectra,ayathesis}.

\section{Data Analysis}
\label{Sec:Analysis}
\subsection{Analysis Method}
\label{Sec:Method}


Our eventual goal is to determine the complete structure of the six-dimensional two-particle correlation for all hadron pair charge combinations. 
Toward this goal the two-particle momentum space was projected onto 2D subspace $(p_{t1},p_{t2})$ by integrating the pseudorapidity and azimuth coordinates $(\eta_1,\eta_2,\phi_1,\phi_2)$ over the detector acceptance for this analysis, $|\eta| \leq 1.3$ and full $2\pi$ azimuth. Projection onto 2D subspace $(p_{t1},p_{t2})$ is achieved by filling 2D binned histograms of the number of pairs of particles for all values of $\eta,\phi$ within the acceptance.
Complementary correlation structures on relative pseudorapidity and azimuth coordinates with integration over transverse momentum acceptance are reported in~\cite{axialci,axialcd}.


The quantities obtained here are ratios of normalized
histograms of {\em sibling} pairs (particles from the same event) to
{\em mixed-event} pairs (each particle of the pair is from a different, but similar event)
in an arbitrary 2D bin with indices $a,b$ representing the values of $p_{t1}$ and $p_{t2}$ (see discussion below).
The normalized pair-number ratio $\hat{r}_{ab}$ introduced in~\cite{isrpp} is here defined by
\bea
\hat{r}_{ab} & \equiv & \hat{n}_{ab,sib}/\hat{n}_{ab,mix},
\label{Eq1}
\eea
where $\hat{n}_{ab,sib} = n_{ab,sib}/\sum_{ab} n_{ab,sib}$ (sum over all 2D
bins), $\hat{n}_{ab,mix} = n_{ab,mix}/\sum_{ab} n_{ab,mix}$, and $n_{ab,sib}$ and
$n_{ab,mix}$ are the inclusive number of sibling and mixed-event pairs, respectively, in 2D bin $a,b$. Histograms and ratios $\hat{r}_{ab}$ were constructed for each charge-sign combination: $(+,+)$, $(-,-)$, $(+,-)$ and $(-,+)$.
Ratio $\hat{r}_{ab}$ is approximately 1, while difference $(\hat{r}_{ab} - 1)$ measures correlation amplitudes and is the quantity reported here.

The exponential decrease in particle yield with increasing $p_t$ degrades the statistical accuracy of $\hat{r}_{ab}$ at larger transverse momentum, thus obscuring the statistically significant correlation structures there.  In order to achieve approximately uniform statistical accuracy across the full $p_t$ domain considered here, nonuniform bin sizes on $p_t$ were used. This was done by noting that the charged hadron $p_t$ distribution, $dN/p_tdp_t$, for Au-Au collisions at $\sqrt{s_{NN}} = 130$~GeV is approximately exponential for $0.15 \leq p_t \leq 2$~GeV/$c$~\cite{spectra} and by dividing the running integral of that exponential distribution into equal bin sizes.  This procedure provides a convenient mapping from $p_t$ to function $X(p_t) \equiv 1 - \exp \{-(m_t - m_0)/0.4~{\rm GeV}\}$ where $0 \leq X \leq 1$,  $m_t = \sqrt{p_t^2 + m_0^2}$, and $m_0$ (here assumed to be the pion mass $m_{\pi}$) is a mapping parameter from coordinate $p_t$ to $X$~\cite{Xmap}.  Equal bin sizes in $X$ therefore have approximately the same number of sibling pairs. For this analysis 25 equal width bins on $X$ from $X(p_t = 0.15~{\rm GeV}/c) = 0.15$ to $X(p_t = 2.0~{\rm GeV}/c) = 0.99$ were used~\cite{ytxyt}.

Normalized pair-number ratios were formed from subsets of events with similar centrality (multiplicities differ by $\leq 100$, except $\leq 50$ for the most-central event class) and primary-vertex location (within 7.5~cm along the beam axis) and combined as weighted (by sibling pair number) averages within each centrality class.
The normalized pair-number ratios for each charge-sign
were combined to form like-sign (LS: $++,--$) and
unlike-sign (US: $+-,-+$) quantities.  The final
correlations reported here were averaged over all four charge-sign quantities,
resulting in the
correlation structures common to all charge-sign combinations.
Hence we refer to these final results as {\em charge-independent} (CI = LS + US) correlations even though they are constructed from quantities which depend on
the charge signs of the hadron pairs. The correlation measure reported here is therefore the CI combination for $\hat r[X(p_{t1}),X(p_{t2})]-1$.


Deviations of event-wise $\langle p_t \rangle$ fluctuations from a central-limit-theorem reference~\cite{cltps,handbook} are measured by scale-dependent ({\em i.e.,} $\eta,\phi$ bin sizes) variance difference $\Delta\sigma^{2}_{p_t:n}$ introduced in \cite{meanptprl}, where it was evaluated at the STAR ($\eta,\phi$) detector acceptance scale. $\Delta\sigma^{2}_{p_t:n}$ can be expressed as a {\em weighted integral} on $(p_{t1},p_{t2})$ of pair-density difference $\rho_{sib} - \rho_{mix}$, where two-particle densities $\rho_{sib}$ and $\rho_{mix}$ are approximated by the event-averaged number of sibling and mixed-event pairs per 2D bin, respectively. Both densities are normalized to the event-averaged total number of pairs.  $\Delta\sigma^{2}_{p_t:n}$ can be rewritten exactly as a discrete sum over $p_t$ products~\cite{cltps} [first line in Eq.~(\ref{Eq2}) below], and the summations approximated in turn by the weighted integral of the pair density difference [second line in Eq.~(\ref{Eq2})] according to,
\bea \label{Eq2}
\Delta\sigma^{2}_{p_t:n} & \equiv &
 \frac{1}{\bar N} \frac{1}{\epsilon} \sum_{j=1}^\epsilon
\sum_{i \neq i'=1}^{N_j} \left(p_{tji} p_{tji'} -
 \hat p^2_t\right) \! \\ \nonumber
& \approx & \frac{1}{\bar N} \int \! \int 
dp_{t1} dp_{t2} p_{t1} p_{t2} \left(\rho_{sib} - \rho_{mix}\right) \\ \nonumber
 & \equiv & \hat{p}^2_t \, \bar{N}  \langle r(p_{t1},p_{t2})-1 \rangle,
\eea
where weighted average $\langle r(p_{t1},p_{t2})-1 \rangle$ is defined in the last line with weight $p_{t1} p_{t2} \rho_{mix}(p_{t1},p_{t2})$ and
the integral of $\rho_{mix}$, $\int \!\! \int dp_{t1} dp_{t2} p_{t1} p_{t2} \rho_{mix}$ is $\bar N^2\, \hat{p}^2_t $~\cite{rhomix}. In Eq.~(\ref{Eq2}) $N_j$ is the event-wise number of accepted particles, $\bar{N}$ is the mean of $N_j$ in the centrality bin, $\epsilon$ is the number of events, $j$ the event index, $\hat p_t$ is the mean of the ensemble-average $p_t$ distribution (all accepted particles from all events in a centrality bin), and $i,i'$ are particle indices. Eq.~(\ref{Eq2}) relates nonstatistical $\langle p_t \rangle$ fluctuations at the acceptance scale to the weighted integral of $\rho_{sib} - \rho_{mix}$, the latter difference being related to the two-particle number correlation density. In the present analysis we measure normalized pair-ratio distributions $\hat r[X(p_{t1}),X(p_{t2})]$ exhibiting two-particle number correlations on $p_t$ which correspond to excess $\langle p_t \rangle$ fluctuations. 

\subsection{Corrections and centrality}
\label{Sec:Corrections}

Corrections were applied to ratio $\hat r$ for two-particle reconstruction inefficiencies due to overlapping space points in the TPC (two trajectories merged into one reconstructed track) and intersecting trajectories which cross paths within the TPC and are reconstructed as more than two tracks.  These corrections were implemented using two-track proximity cuts~\cite{trackcuts} at various radial positions in the TPC in both the longitudinal (drift) and transverse directions (approximately along the pad rows).  The track pair cuts were applied to both $\rho_{sib}$ and $\rho_{mix}$ as in HBT analyses~\cite{starhbt}.
Small-momentum-scale correlation structures due to quantum interference,
Coulomb and strong final-state interactions~\cite{starhbt} were suppressed by eliminating sibling and mixed-event track pairs ($\sim$3\% of total pairs) with $|\eta_1 - \eta_2| < 0.3$, $|\phi_1 - \phi_2| < \pi/6$ (azimuth), $|p_{t1} - p_{t2}| < 0.15$~GeV/$c$, if $p_t < 0.8$~GeV/$c$ for either particle.  The small-momentum-scale correlation (SSC) structures are most prominent in the lower-$p_t$ domain of the 2D $(p_{t1},p_{t2})$ space along the $p_{t1} = p_{t2}$ diagonal and were shown to be similar in amplitude and location to simulations~\cite{mevsim} which account for quantum interference correlations and Coulomb final-state interaction effects using pair weights determined by HBT analyses for these data~\cite{starhbt}.  The preceding cuts were optimized~\cite{ayathesis} to eliminate SSC structure without affecting the large-momentum-scale correlation (LSC) structure which is of primary interest here. The track-pair cuts generally have small effects on the LSC; uncertainties which result from application of these cuts are discussed in Sec.~\ref{Sec:Errors} and are negligible compared to the large momentum scale structures studied here.

Four centrality classes labeled (a) - (d) for central to peripheral were defined by cuts on TPC track multiplicity $N$ within the acceptance
by (d) $0.03 < N/N_0 \leq 0.21$, (c) $0.21 < N/N_0 \leq 0.56$, (b) $0.56 < N/N_0 \leq 0.79$ and (a) $N/N_0 > 0.79$, corresponding respectively to approximate fraction of total cross section ranges 40-70\%, 17-40\%, 5-17\% and 0-5\%.
$N_0$ is the end-point~\cite{endpoint} of the minimum-bias multiplicity distribution.

\begin{figure}[h]
\includegraphics[keepaspectratio,width=1.65in]{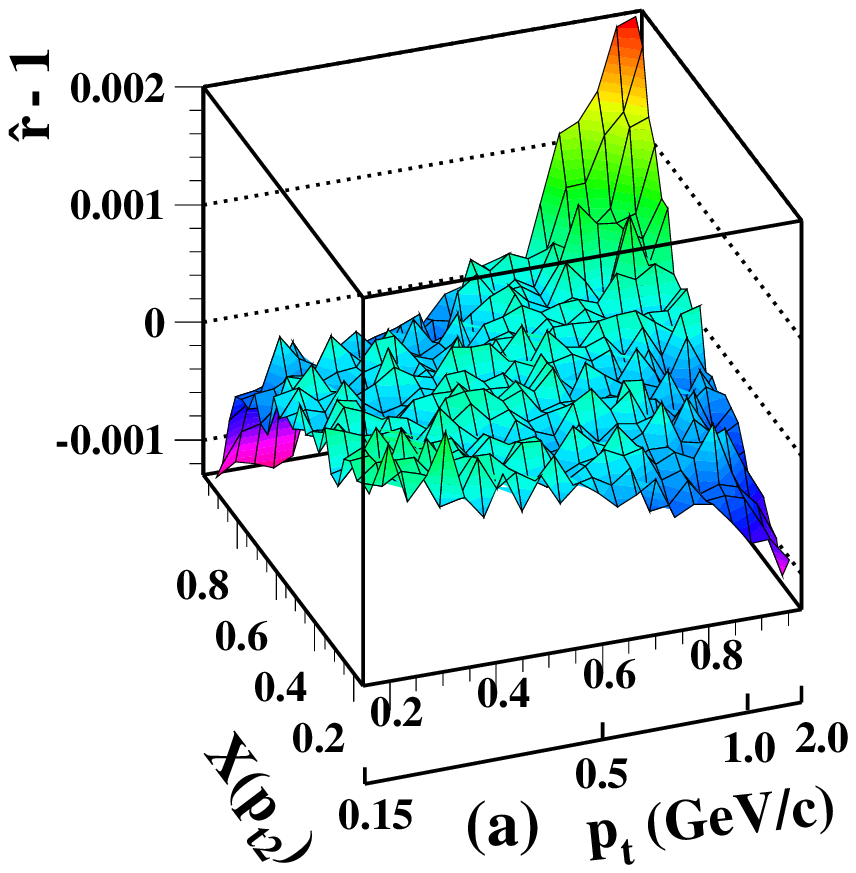}
\includegraphics[keepaspectratio,width=1.65in]{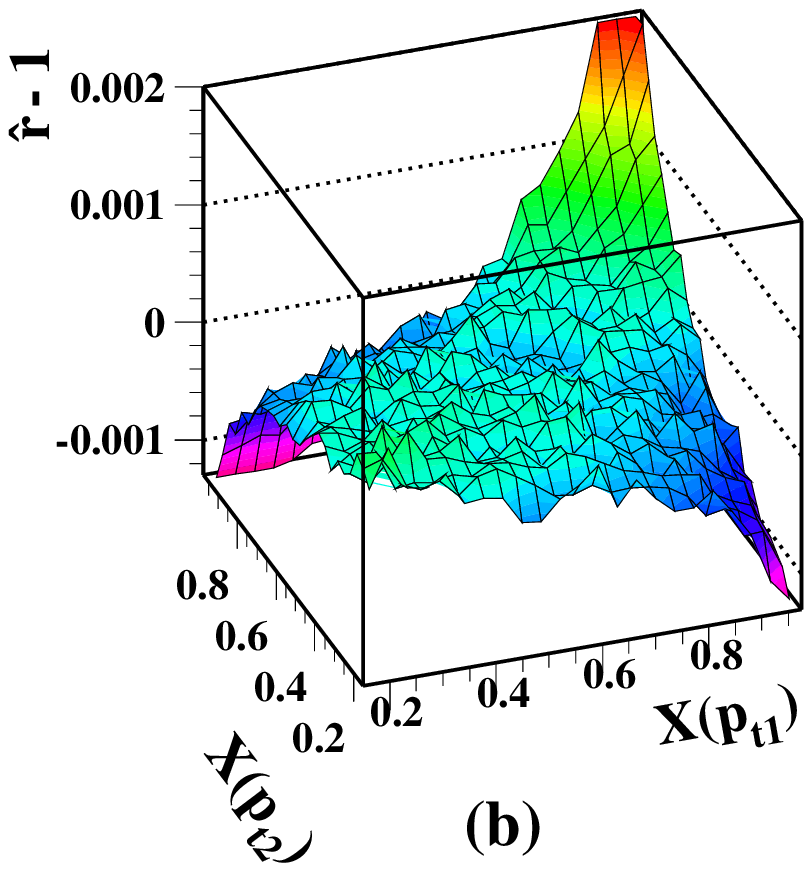}
\includegraphics[keepaspectratio,width=1.65in]{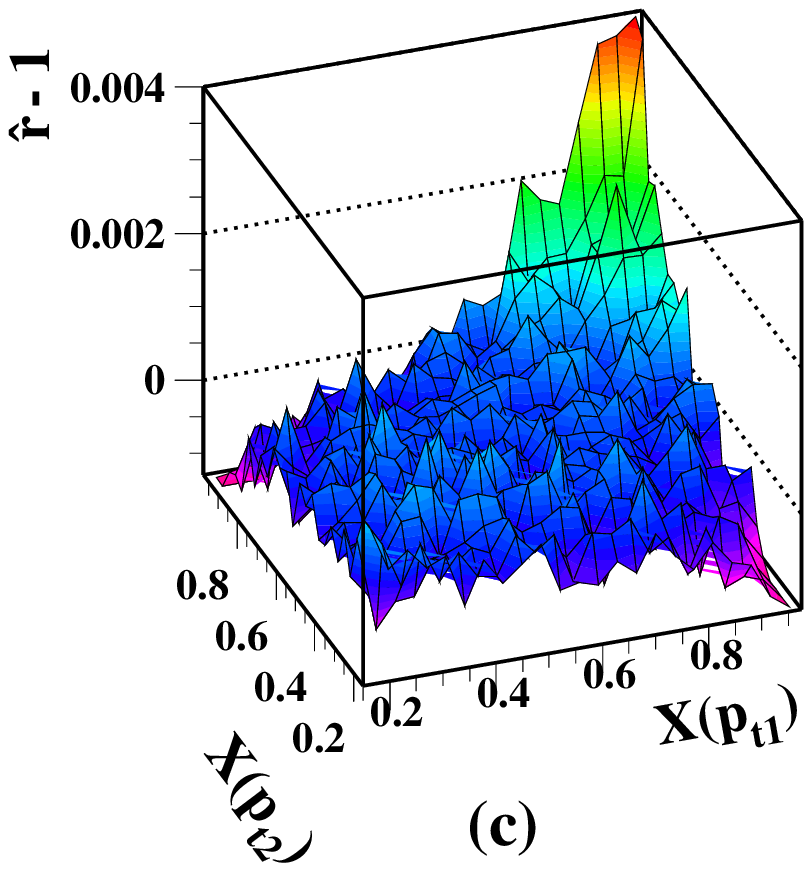}
\includegraphics[keepaspectratio,width=1.65in]{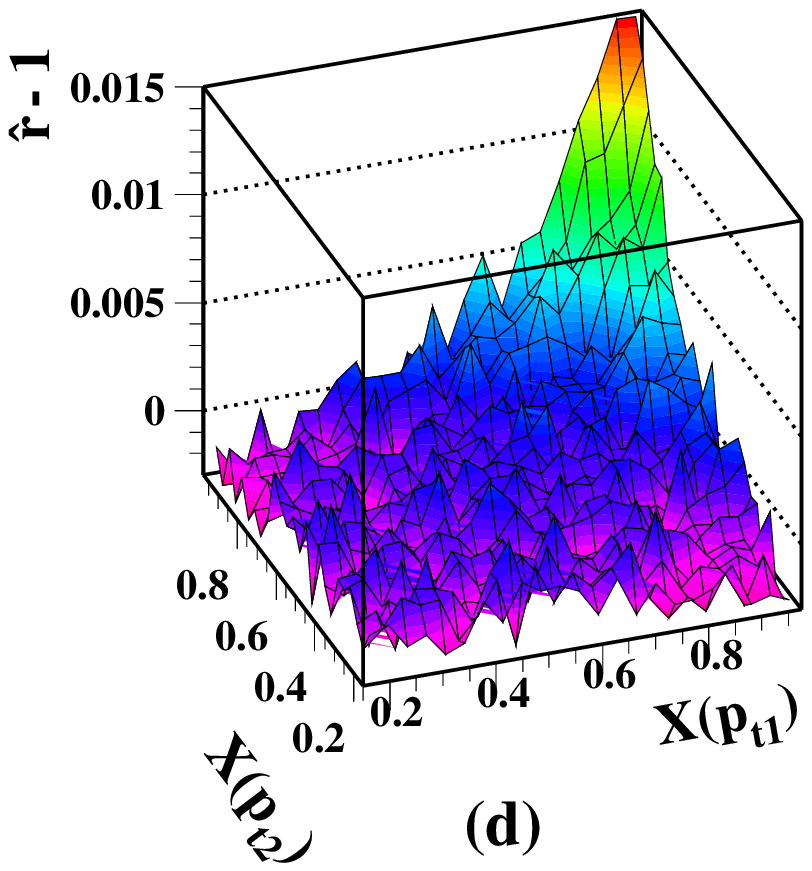}
\caption{\label{Figure1} Symmetrized pair-density net ratios $\hat r[X(p_{t1}),X(p_{t2})] - 1$ for all nonidentified charged primary particles for (a) most-central, (b) mid-central, (c) mid-peripheral, and (d) peripheral Au-Au collision events at $\sqrt{s_{NN}} = 130$~GeV/$c$.  Note the scale change for panels (c) and (d) and auxiliary $p_t$ scale in units GeV/$c$ in panel (a). SSC were removed using track pair cuts (see text). Errors are discussed in Sec.~\ref{Sec:Errors}.}
\end{figure}


The centrality dependence of quantity $\hat r - 1$ is shown in Fig.~\ref{Figure1} as perspective views for the four centrality classes used here. This correlation measure represents the number of correlated particle pairs per final-state pair in each 2D bin, and therefore contains a dilution factor $1/\bar N$ relative to the LSC measure presented in~\cite{axialcd}, $\bar N (\hat r - 1)$ whose amplitudes are of order one.
The structures in Fig.~\ref{Figure1} are therefore numerically a few {\em permil} for central Au-Au collisions but are highly significant statistically as seen by comparing to the statistical errors.
The dominant features in Fig.~\ref{Figure1} are 1) a large-momentum-scale correlation `saddle' structure with positive curvature along the $X(p_t)_{\Sigma} \equiv X(p_{t1})$ $+$ $X(p_{t2})$ sum direction from $[X(p_{t1}),X(p_{t2})]$ (0,0) to (1,1) and a corresponding negative curvature along the $X(p_t)_{\Delta} \equiv X(p_{t1})$ $-$ $X(p_{t2})$ difference direction from $[X(p_{t1}),X(p_{t2})]$ (0,1) to (1,0), and 2) a narrow peak structure at large $X(p_t)$ ($p_t > 0.6$~GeV/$c$). With increasing centrality the negative curvature of the LSC saddle shape along the difference variable increases in magnitude, the positive curvature along the sum variable decreases, and the magnitude of the peak at large $X(p_t)$ also decreases. Without the SSC cuts a relatively small peaked structure with amplitude of order 0.004 (peripheral) to 0.0005 (central) is present for $X(p_t) < 0.3$ ($p_t < 0.25$~GeV/$c$) which weakens in amplitude but visibly persists to $X(p_t) < 0.6$ ($p_t < 0.5$~GeV/$c$).

An upper limit estimate for resonance contributions was obtained using Monte Carlo simulations~\cite{mevsim} assuming 70\% of the primary charged particle production is from resonance decays. The correlations were simulated by populating the events with a sufficient number of $\rho^0 , \omega$ two-body decays to account for 70\% of the observed multiplicity. These two-body decay processes produced a small saddle-shape correlation with curvature opposite to the data and amplitude at the corners approximately 0.0002 for the most-central data, increasing as $1/\bar{N}$ for the remaining centrality bins. The saddle-shape structures in Fig.~\ref{Figure1} cannot be explained with resonance decays.

The same analysis applied to Pb-Pb collisions in $1.1 < y_{cm} < 2.6$ at the CERN SPS did not reveal any statistically significant CI correlations~\cite{jeffdata} when SSC (see Sec.~\ref{Sec:Corrections}) were removed with pair cuts. The analysis in \cite{anticic} of proton + proton and various nucleus + nucleus collision data from the CERN SPS for $1.1 < y_{cm} < 2.6$ without those pair cuts revealed SSC peaks at low $X(p_t)$ along the $X(p_{t})_{\Sigma}$ direction.

\subsection{Error analysis}
\label{Sec:Errors}

Per-bin statistical errors for $\hat r - 1$ in Fig.~\ref{Figure1}
range from $\sim$6-9\% of the maximum correlation amplitude for each centrality
[typically 0.00015, 0.00011, 0.00035 and 0.001 for centralities (a)-(d) respectively] and are approximately uniform, by design, over the 2D domain on $X(p_t)$. Statistical errors for $\bar N (\hat r - 1)$ ($\sim$0.1 - 0.15) are less dependent on centrality.

Systematic errors were estimated as in \cite{axialcd,meanptprl} and
are dominated by the 7\% non-primary background contamination~\cite{spectra} whose correlation with primary particles is unknown.  The upper limit on the systematic error from this source was estimated by assuming the number of correlated pairs associated with background-primary pairs of particles could range from zero up to the amount which would occur among 7\% of the primary particles and the remaining primaries. This conservative assumption produces an overall $\pm$7\% uncertainty relative to the correlation amplitudes in Fig.~\ref{Figure1} throughout the domain for $X(p_{t1,2}) > 0.4$.  This error increases to $\pm$16\% at lower $X(p_t)$ where the contamination fraction is larger and is about $\pm$12\% in the off-diagonal corners of the $[X(p_{t1}),X(p_{t2})]$ domain. Multiplicative factors for quantity $\hat r - 1$ which correct for the non-primary background contamination range from 1.0, assuming background-primary particle pairs are correlated and increase both $\hat n_{ab,sib}$ and $\hat n_{ab,mix}$ by $2\times7$\%~=~14\%, to 1.14 if background-primary particle pairs are uncorrelated but the non-primary background contributes 14\% to $\hat n_{ab,mix}$. Multiplication of the data in Fig.~\ref{Figure1} by average factor 1.07 provides an estimate of the background corrected correlation amplitudes.

Additional sources of systematic error were evaluated.
Uncertainty in the two-track inefficency corrections have modest effects along the $X(p_{t1}) = X(p_{t2})$ diagonal ($<2$\%) and are negligible elsewhere. Tracking anomalies caused when particle trajectories intersect the TPC high-voltage central membrane significantly affect the $X(p_{t1,2}) < 0.2$ domain corresponding to the single bin at lowest $X(p_t)$, and the diagonal bins for $X(p_{t1,2}) < 0.4$ by 20\%.
Final multiplicative correction estimates (not applied in Fig.~\ref{Figure1}) and total systematic errors for $\hat r - 1$ varied respectively from 1.07 and $\pm$7\% for $X(p_{t1,2}) > 0.4$ up to 1.16 and $\pm$16-20\% for $X(p_{t1,2}) < 0.4$ and 1.12 and $\pm$12\% in the off-diagonal corners [{\em i.e.,} near (0,1) and (1,0)].

Other potential sources of systematic error were studied and determined to have
negligible effects including primary vertex position uncertainty perpendicular to the beam direction, variation of tracking acceptance and efficiency with primary vertex location along the axis of the TPC, TPC drift speed and/or timing-offset fluctuation, sporadic outages of TPC read-out electronic components, angular
resolution, multiplicity and primary vertex position bin sizes used for producing mixed events, and charge sign dependence of the tracking efficiency.
Conversion electron contamination is suppressed by the lower $p_t$ acceptance cut and also by the pair cuts described in Sec.~\ref{Sec:Corrections} and also makes negligible contribution to the systematic error.


\section{Modeling One- and Two-Particle Distributions on $p_t$}
\label{Sec:Dist}

Two features dominate the data in Fig.~\ref{Figure1}: 1) a large-momentum
scale saddle shape and 2) a peak at large $X(p_{t1})$ and $X(p_{t2})$. In
this section results from Monte Carlo collision models are
analyzed in order to gain insight into the dynamical origin(s) of these two
correlation structures in the data.  Based on this study an
analytical function is obtained which accurately describes the saddle shape
and in Sec.~\ref{Sec:Model} this function is used to fit the 2D correlation
data.
 
\begin{figure}[h]
\includegraphics[keepaspectratio,width=1.65in]{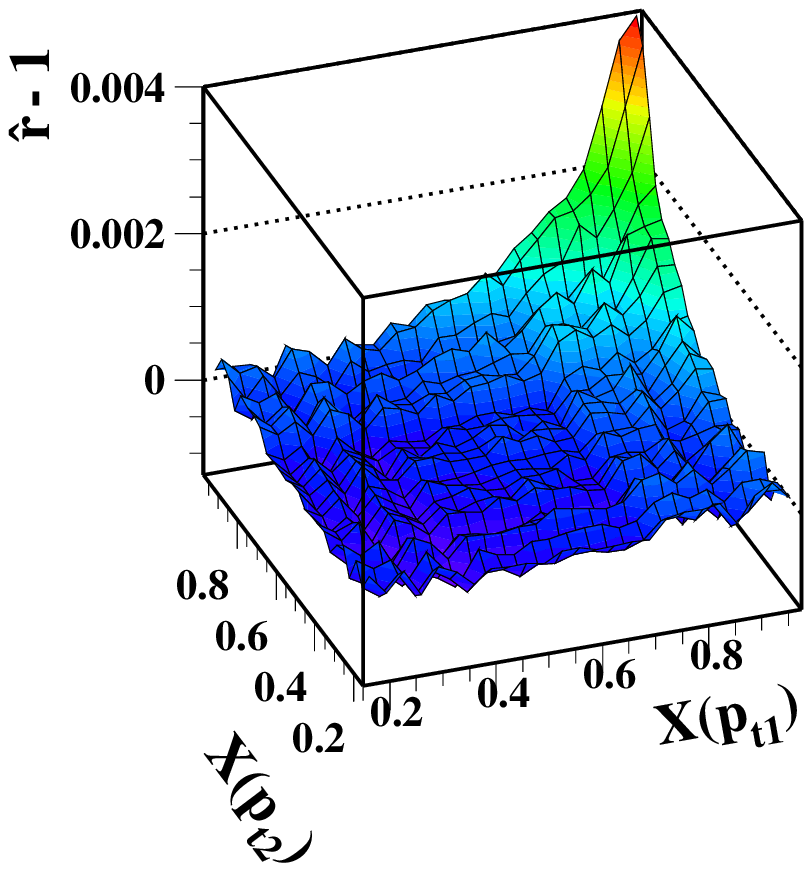}
\includegraphics[keepaspectratio,width=1.65in]{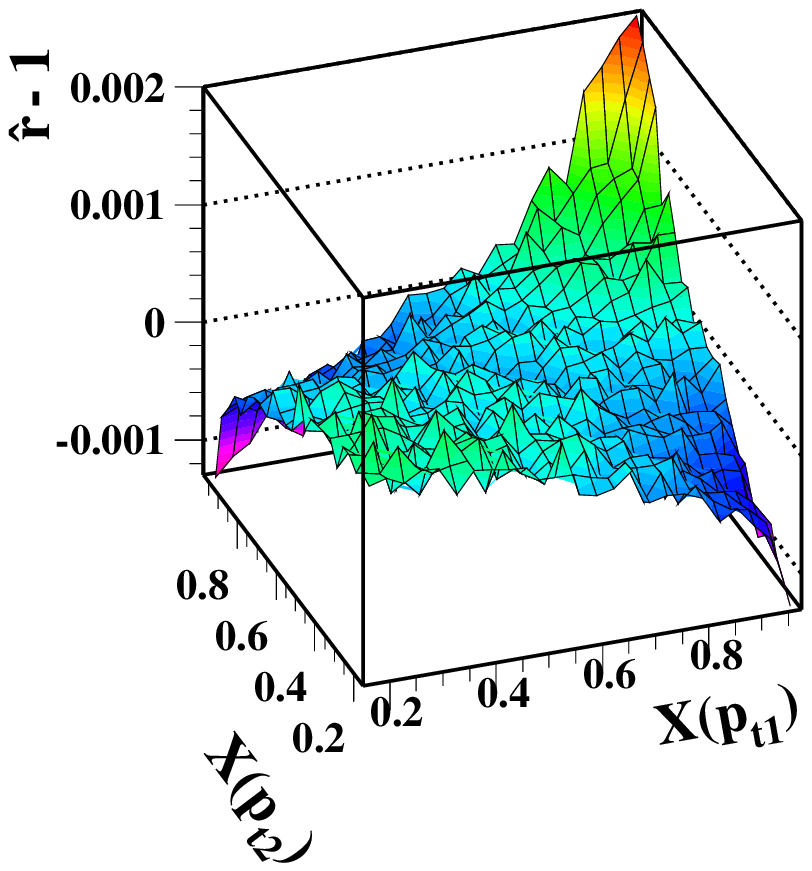}
\caption{\label{Figure2} Symmetrized pair-density ratio $\hat r[X(p_{t1}),X(p_{t2})] - 1$ for unidentified charged particles and for central Au-Au collisions. Left panel: Default {\sc Hijing}~\cite{jetquench} with jet quenching,
Right panel: a Monte Carlo model~\cite{mevsim} which simulates event-wise global temperature fluctuations (see text).}
\end{figure}

The high energy nuclear collision model {\sc hijing}~\cite{jetquench}, which includes longitudinal color-string fragmentation and perturbative quantum chromodynamics (pQCD) based jet production and jet quenching, exhibits significant correlation structure at higher $p_t$ [$X(p_{t1}) + X(p_{t2}) > 1.6$] as shown in the left-hand panel of Fig.~\ref{Figure2} for central Au-Au collisions. The predictions, which include jet production with jet quenching (default parameters) are qualitatively different from the data in Fig.~\ref{Figure1}, failing to produce any saddle-shape, but suggest the type of correlation structure produced by jets.
The general structure of the {\sc hijing} predictions suggests that the peaks in the data at higher $X(p_t)$ are at least partly due to initial-state partonic scattering and fragmentation. Other theoretical models which combine initial-state parton scattering, energy loss, dissipation, rescattering and recombination~\cite{ampt,reco} may eventually explain these correlation data, but relevant predictions are not available at this time.

The saddle-shape correlation spans the entire momentum scale studied here, suggesting event-wise fluctuations of global event characteristics ({\em e.g.} temperature and/or collective velocity of the bulk medium) as a possible source. If heavy ion collisions at RHIC thermalize then an ensemble of collision events would be characterized by a distribution of event-wise equilibrium temperatures reflecting event-to-event fluctuations in the initial conditions and time evolution of each colliding system. Based on this idea the transverse momentum correlations for an ensemble of such events can be predicted using a Monte Carlo model in which charged particle production is generated by sampling the inclusive single-particle $(p_t,\eta,\phi)$ distribution obtained from the data.  At mid-rapidity the inclusive distribution on $p_t$ for $0.15 \leq p_t \leq 2$~GeV/$c$ is well approximated by $\exp (-m_t/T) \equiv \exp (-\beta m_t)$~\cite{spectra} where $T$ is an effective temperature~\cite{ssh} or inverse slope parameter and $\beta = 1/T$.  Events were generated by sampling $\exp (-m_t/T)$ where $T$ fluctuates randomly from event-to-event according to a gaussian distribution about mean value $T_0 = 1/\beta_0$; $T_0$ was determined by the measured $p_t$ spectrum.

The result of this Monte Carlo model for central Au-Au collisions at 130~GeV is shown in the right-hand panel of Fig.~\ref{Figure2} where the mean and standard deviation (gaussian sigma) of the event-wise temperature distribution are $T_0 = 200$~MeV and $\sigma_T/T_0 = 1.5$\%.  The predicted correlations are not sensitive to $T_0$ but the overall correlation amplitude is directly sensitive to $\sigma_T/T_0$ which was adjusted to approximate the overall amplitude of the data in Fig.~\ref{Figure1}(a).  The global temperature fluctuation model accurately describes the saddle-shape. An analytical function based on this approach is derived in the remainder of this section and is used in the following section (Sec.~\ref{Sec:Model}) to fit the data.

We seek an analytical representation of the LSC saddle-shape
structure of the data in Fig.~\ref{Figure1} that is both mathematically
compact and physically motivated in order to conveniently characterize the centrality dependence and to infer thermodynamic properties of the medium. The above Monte Carlo results indicate that a successful representation should involve an averaging over the inverse slope parameter. In general the inverse temperature $\beta$ can vary from event-to-event as well as internally within each event, reflecting the possibility
of relative ``hot spots'' and ``cold spots'' in the final-state particle distributions. The number, location in source coordinates ({\em e.g.,} $\eta_z$ $-$ space-time rapidity~\cite{ssh} and $\varphi$ $-$ azimuth), amplitude, and angular extent of these perturbations in $\beta$ may vary for each event. In addition, for realistic collision systems both thermal and collective motions are involved such that parameter $\beta$ becomes an inverse {\em effective} temperature~\cite{ssh} where fluctuations in $\beta$ could result from fluctuations in the local temperature of the flowing medium, the collective flow velocity itself, or a combination of both effects~\cite{tempflow}. Event-wise effective temperature is therefore represented by distribution $T(\eta_z,\varphi)$ on source coordinates $\eta_z$ and $\varphi$, and similarly for $\beta(\eta_z,\varphi)$.

\begin{figure}[h]
\includegraphics[keepaspectratio,width=1.6in]{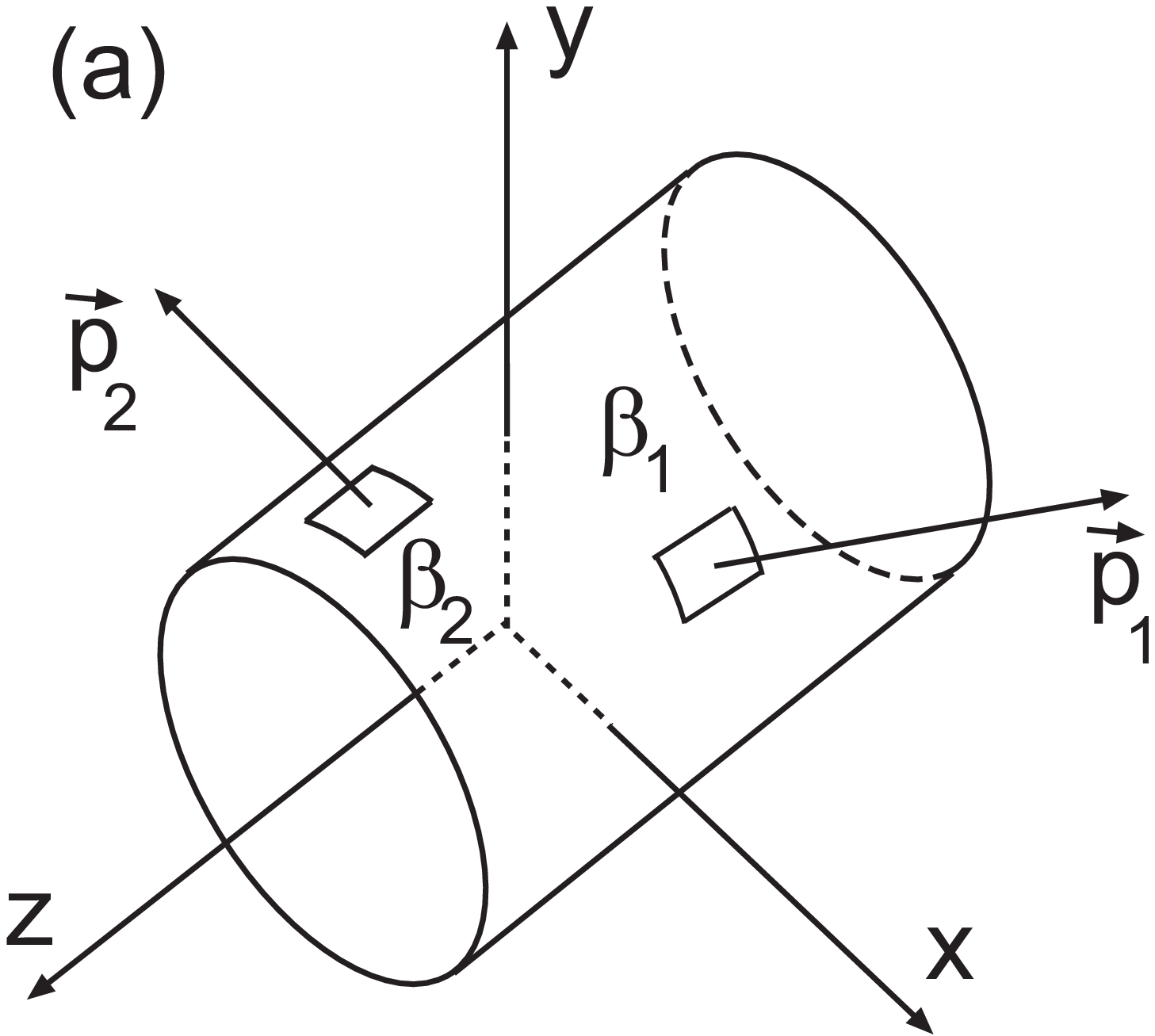}
\includegraphics[keepaspectratio,width=1.6in]{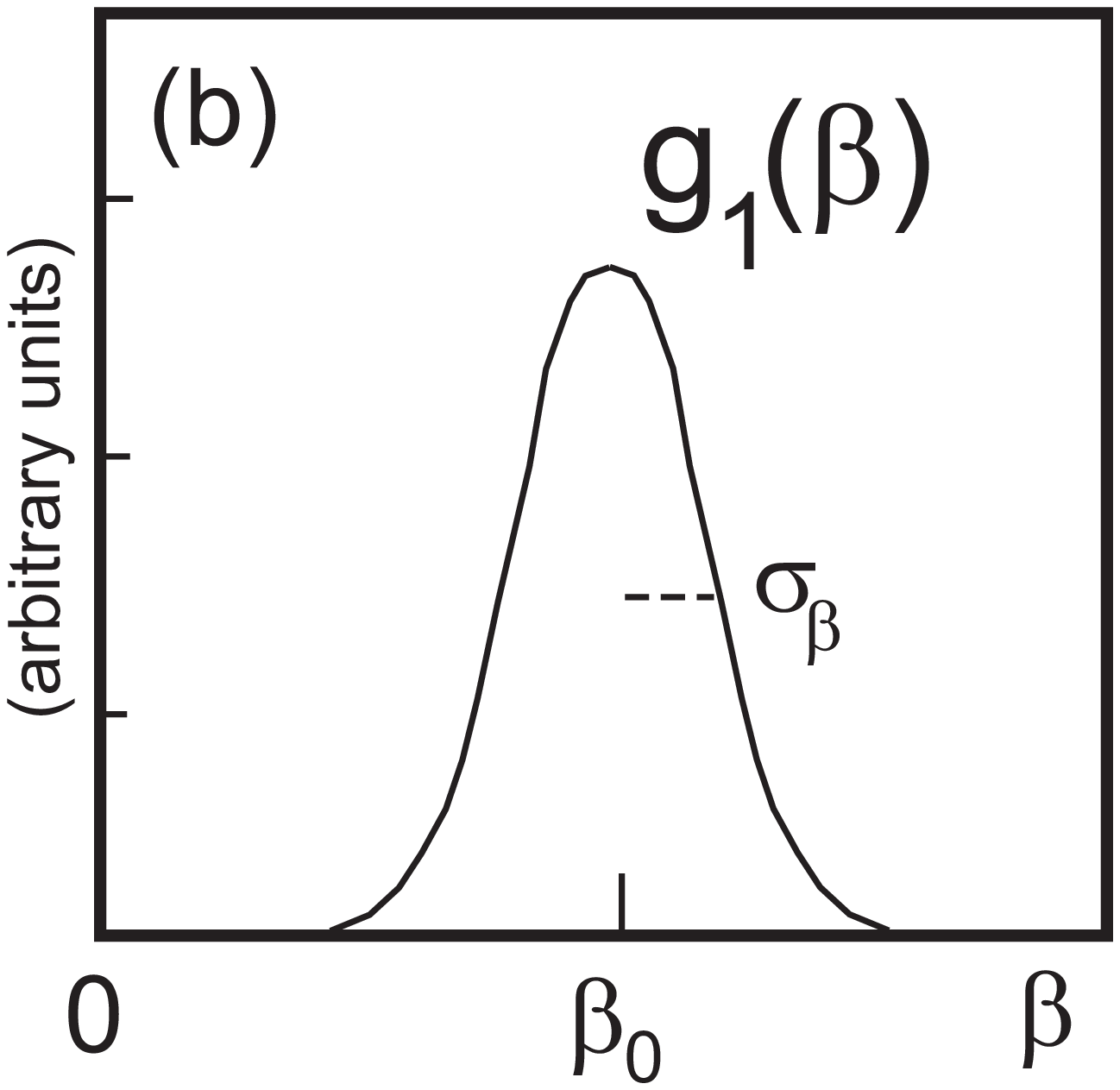}
\includegraphics[keepaspectratio,width=1.6in]{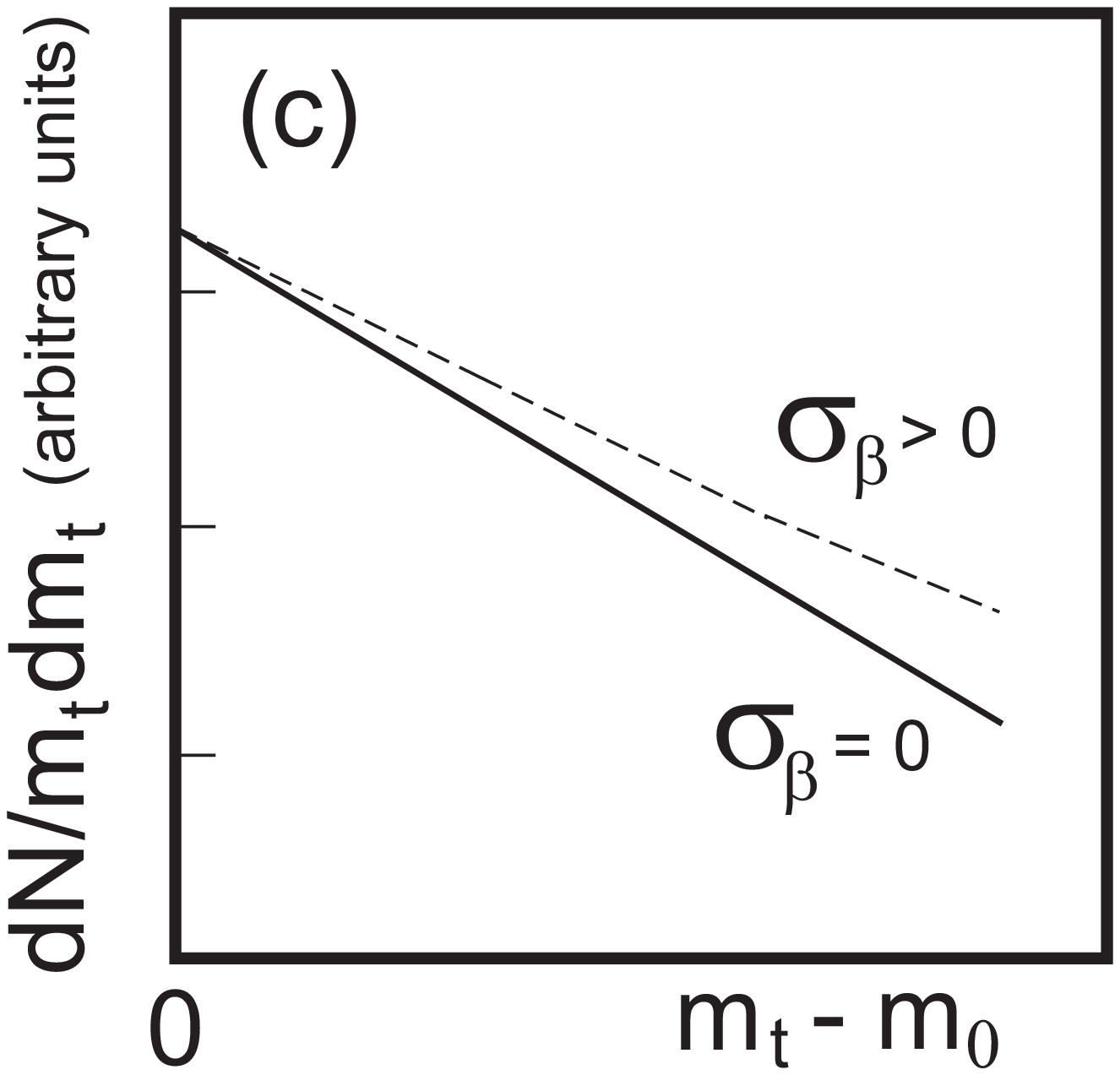}
\includegraphics[keepaspectratio,width=1.6in]{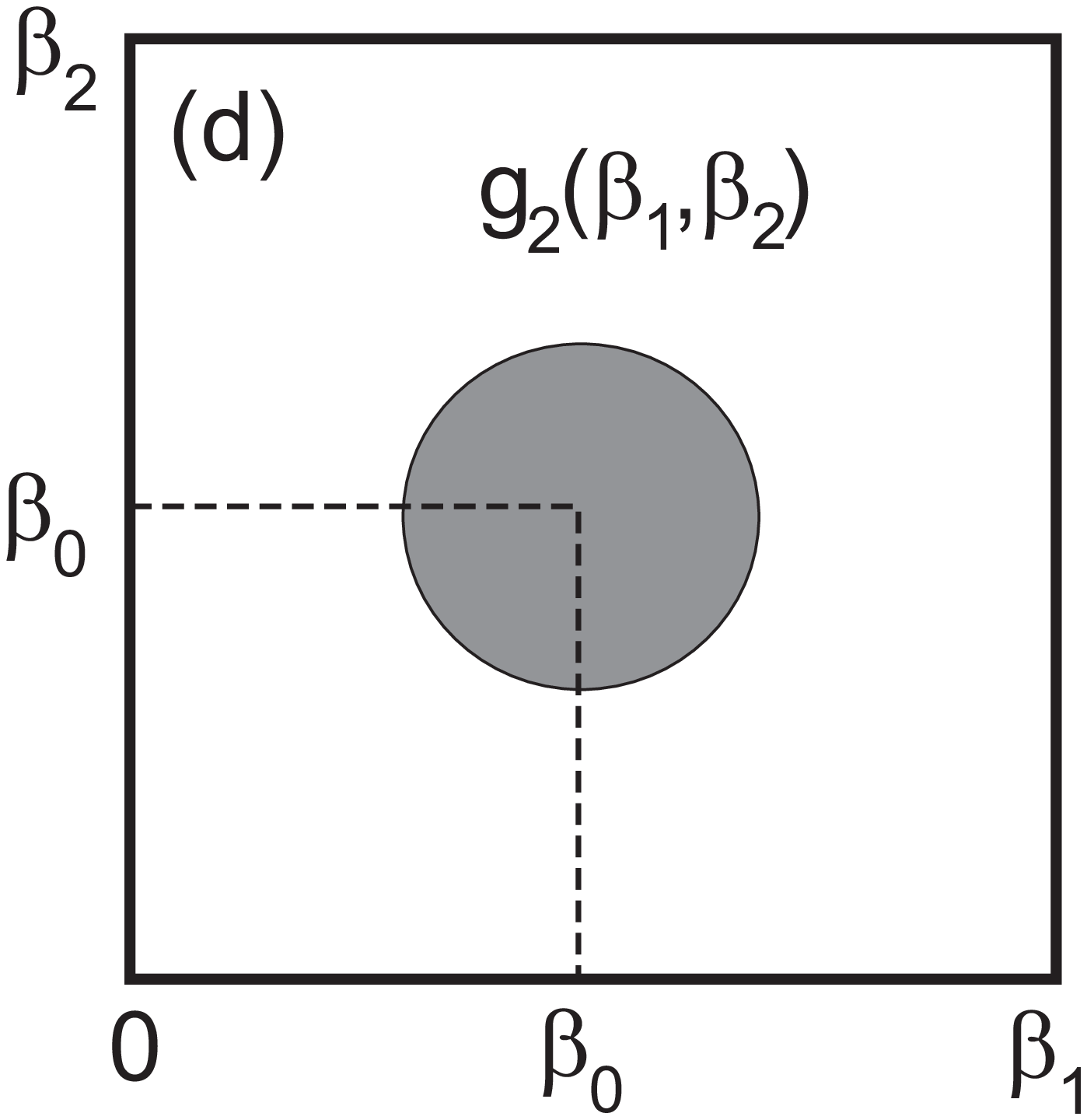}
\includegraphics[keepaspectratio,width=1.6in]{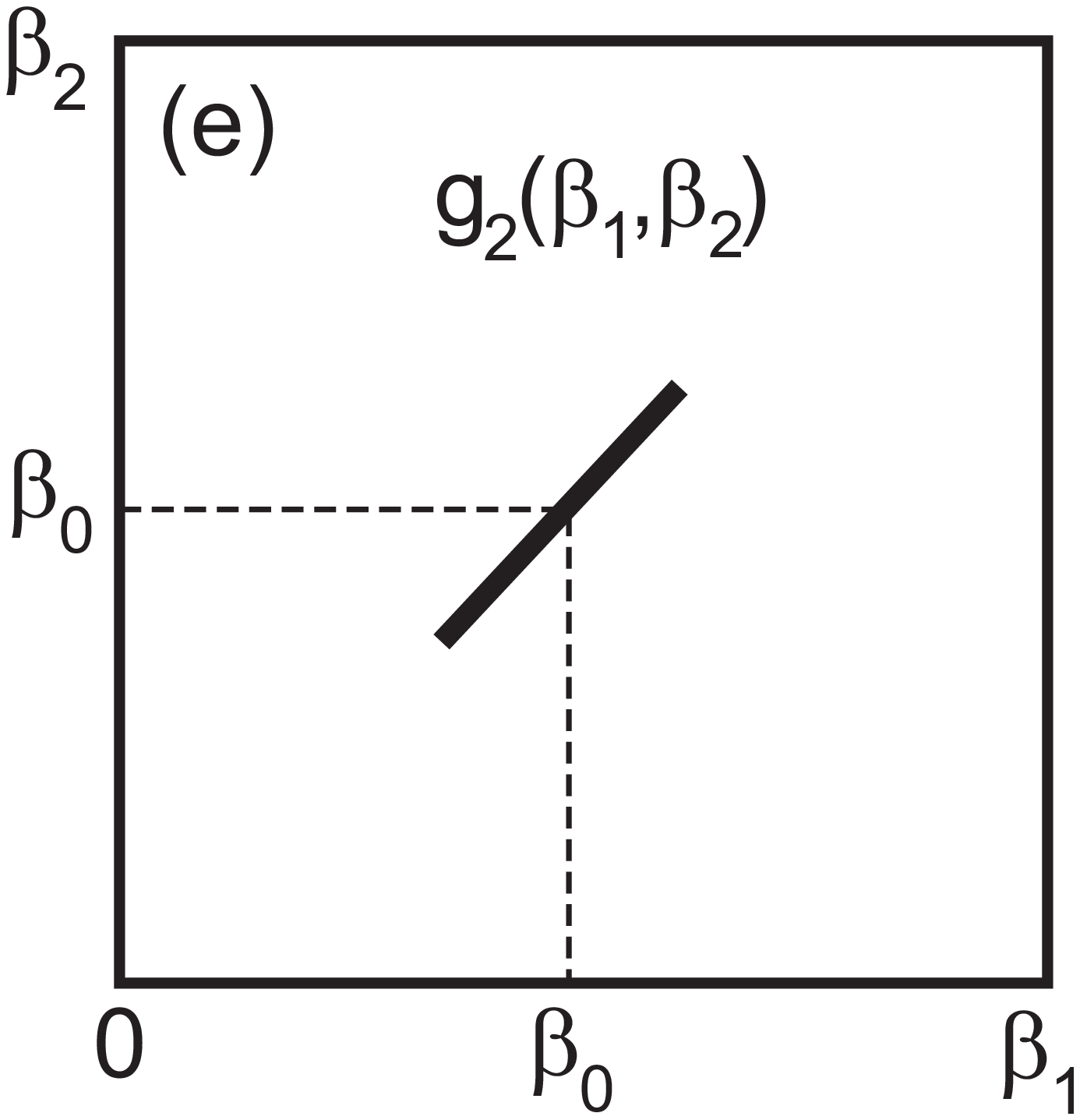}
\includegraphics[keepaspectratio,width=1.6in]{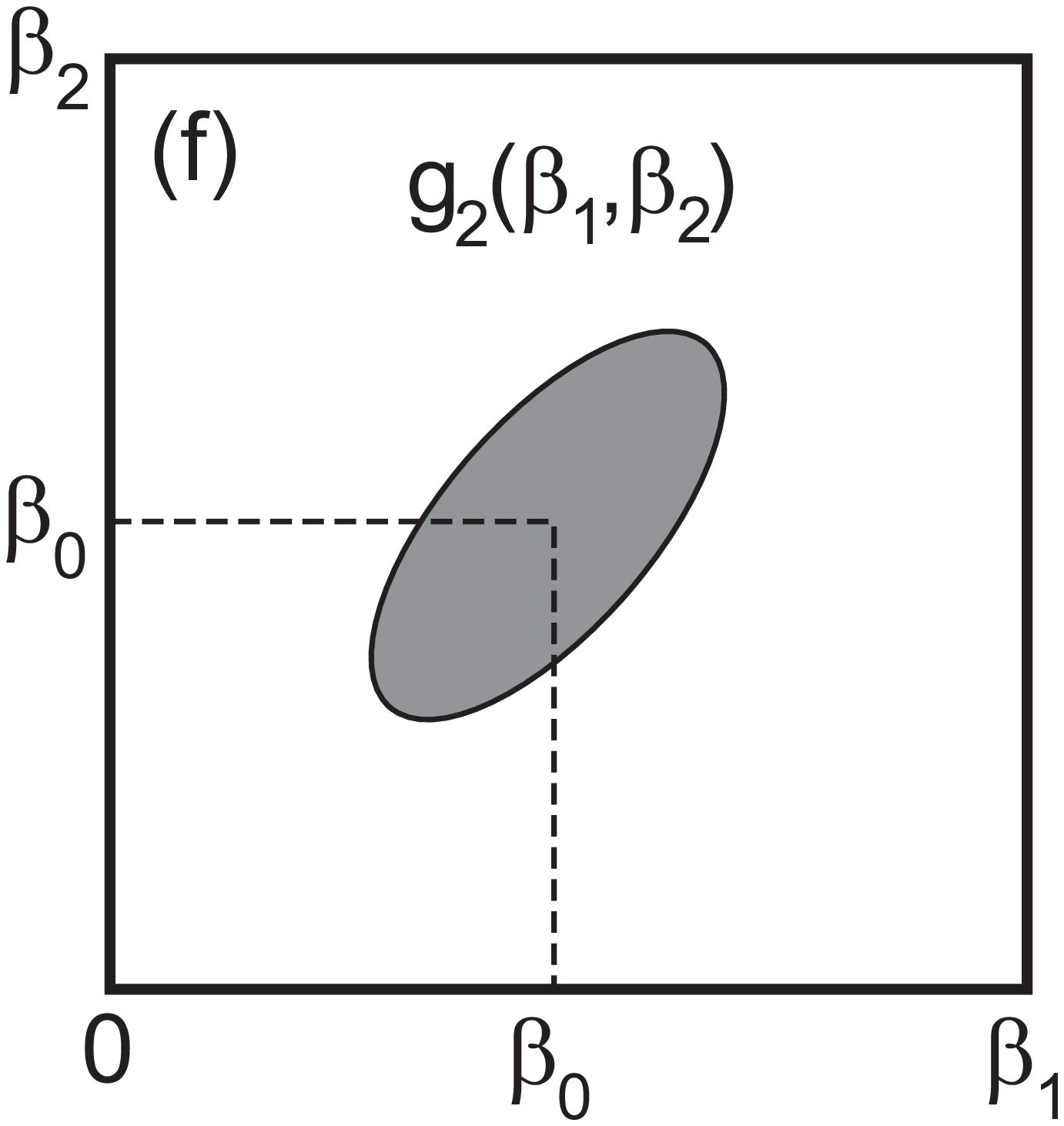}
\caption{\label{Figure3}
Diagrams illustrating the temperature fluctuation model.  Panel (a): Source coordinates with two final state particles sampling local inverse temperatures $\beta_1 = \beta(\eta_{z1},\varphi_1)$ and $\beta_2 = \beta(\eta_{z2},\varphi_2)$. Panel (b): Distribution $g_1(\beta)$ of sampled $\beta$ values for all particles in all events of a centrality bin with mean $\beta_0$ and standard deviation $\sigma_\beta$.  Panel (c): Thermal model inclusive charged particle yield $dN/m_tdm_t$ at mid-rapidity versus $m_t - m_0$ with no temperature fluctuations ($\sigma_\beta = 0$, solid line) and with temperature fluctuations ($\sigma_\beta > 0$, dashed line). Panel (d): 2D distribution, $g_2(\beta_1,\beta_2)$, of sampled pairs $\beta(\eta_{z1},\varphi_1)$ and $\beta(\eta_{z2},\varphi_2)$ when there are no point-to-point temperature correlations within each source but large temperature variations within each event (non-equilibrium sources).  Panel (e): Same as (d) except for global temperature fluctuations where each event is equilibrated but the equilibrium temperature fluctuates from event-to-event.  Panel (f): Same as (d) except point-to-point temperature correlations occur within each event as evidenced by the positive covariance of distribution $g_2(\beta_1,\beta_2)$.}
\end{figure}

The momentum of a particle at $\eta_z,\varphi$ in the final stage of the collision system is obtained by sampling thermal distribution $\exp[-m_t/T(\eta_z,\varphi)] = \exp[-\beta(\eta_z,\varphi)m_t]$ as illustrated in the diagram in Fig.~\ref{Figure3}(a). In general the histogram of sampled $T(\eta_z,\varphi)$ or $\beta(\eta_z,\varphi)$ values for all particles in all events in the event ensemble, $g_1(\beta)$, could be like the generic peaked distribution in Fig.~\ref{Figure3}(b) with mean $\beta_0$ and standard deviation $\sigma_\beta$. The inclusive $m_t$ distribution is then obtained by convoluting thermal distribution $\exp[-\beta (m_t - m_0)]$ with $g_1(\beta)$ given by
\bea
\frac{dN}{m_tdm_t} & = & A \int_0^\infty d\beta g_1(\beta) {\rm e}^{-\beta (m_t - m_0)}
\label{Eq:XX}
\eea
where $A$ is a normalization constant. The
global temperature fluctuation model is recovered when $T(\eta_z,\varphi)$ is independent of source coordinate but varies from event-to-event. 

In the Monte Carlo model event-wise $T = 1/\beta$ was obtained by sampling a gaussian distribution.  It is therefore reasonable to represent $g_1(\beta)$ by a peaked distribution which is here assumed to be a gamma distribution~\cite{tann} in order to obtain an analytic solution of the integral in Eq.(\ref{Eq:XX}) given by
\bea
\frac{dN}{m_tdm_t} & = & A/[1 + \beta_0 (m_t - m_0)/n_{fluct}]^{n_{fluct}},
\label{Eq:XX1}
\eea
a L\'evy distribution~\cite{wilk}, where $1/n_{fluct} = \sigma^2_\beta /\beta_0^2$ is the relative variance of $g_1(\beta)$. The finite width of $g_1(\beta)$ produces a net increase in the yield at higher $m_t$ as illustrated in Fig.~\ref{Figure3}(c). We emphasize that any finite-width peaked function $g_1(\beta)$ results in an $m_t$ distribution which decreases less rapidly with increasing $m_t$ than thermal spectrum e$^{-\beta_0 m_t}$. The assumption of a gamma distribution for $g_1(\beta)$ is therefore not essential but is used for mathematical convenience and is justified by the capability of the $m_t$ distribution in Eq.~(\ref{Eq:XX1}) to describe the inclusive data.  We note however that deviations of the measured $m_t$ distribution from a thermal spectrum, quantified by exponent $n$ in the power-law $m_t$ distribution~\cite{spectra}, can result from transverse expansion~\cite{ssh} in addition to local and event-to-event fluctuations in $\beta(\eta_z,\varphi)$ assumed in deriving Eqs.~(\ref{Eq:XX}) and (\ref{Eq:XX1}).  Consequently, fitting the $1/m_t\, dN/ dm_t$ spectra to obtain the power-law exponent $n$ cannot by itself determine the relative variance of the effective temperature distribution, $1/n_{fluct}$, which is related to the degree of equilibration.

Similarly the two-particle distribution on $(m_{t1},m_{t2})$ is obtained by convoluting the two-particle thermal distribution $\exp[-\beta_1(m_{t1} - m_0)] \exp[-\beta_2(m_{t2} - m_0)]$ with the 2D distribution of pairs of inverse effective temperature parameters $(\beta_1,\beta_2)$, where particles 1 and 2 sample local thermal distributions determined by $\beta(\eta_{z1},\varphi_1)$ and $\beta(\eta_{z2},\varphi_2)$, respectively [see Fig.~\ref{Figure3}(a)].  The distribution of $(\beta_1,\beta_2)$ for all pairs of particles used in all events in the ensemble defines a 2D histogram and 2D distribution, $g_2(\beta_1,\beta_2)$, illustrated in Fig.~\ref{Figure3}, panels (d)-(f) for three hypothetical cases.  If the event ensemble distribution on $\beta$ has finite width ($\sigma_\beta > 0$), but are point-to-point uncorrelated within each event, then $g_2(\beta_1,\beta_2)$ is symmetric on $\beta_1$ {\em vs} $\beta_2$ (zero covariance) as shown in Fig.~\ref{Figure3}(d). For uncorrelated $\beta$ fluctuations or for mixed-event pairs, $g_2$ factorizes as $g_2(\beta_1,\beta_2)=g_1(\beta_1) g_1(\beta_2)$, implying zero covariance. On the other hand, if every event is thermally equilibrated, then each pair of particles from a given event samples the same value of $\beta$ where $\beta_1 = \beta_2$. For this case (global temperature fluctuation model) $g_2(\beta_1,\beta_2)$ limits to a diagonal line distribution illustrated in Fig.~\ref{Figure3}(e) and given by $g_2(\beta_1,\beta_2) \propto g^{\prime}_1(\beta_1) \delta(\beta_1 - \beta_2)$, where $\delta(\beta_1 - \beta_2)$ is a Dirac delta-function.  In this case $g_2$ has maximum covariance and represents the conventional picture of an ensemble of equilibrated events with event-wise fluctuations in the global temperature. In general $g_2(\beta_1,\beta_2)$ may have an intermediate covariance as illustrated in Fig.~\ref{Figure3}(f).  In this case if $g_2(\beta_1,\beta_2)$ is expressed as a product of a gamma distribution on the sum direction, $\beta_\Sigma = \beta_1 + \beta_2$ multiplied by a gaussian on $\beta_\Delta = \beta_1 - \beta_2$ (for mathematical convenience), then an analytic expression for the two-particle distribution results, given by a 2D L\'evy distribution
\bea \label{fsib}
F_{sib} & \propto & \left(1 + \frac{\beta_0 m_{t\Sigma}}{2n_{\Sigma}}
\right) ^{-2n_{\Sigma}}
\left[ 1 - \left( \frac{\beta_0 m_{t\Delta}}{2n_{\Delta} + \beta_0 m_{t\Sigma}}
\right) ^2 \right]^{-n_{\Delta}}
\eea
on sum and difference variables $m_{t\Sigma} \equiv m_{t1}
+ m_{t2} - 2m_0$ and $m_{t\Delta} \equiv m_{t1} - m_{t2}$. Inverse exponents $1/n_{\Sigma}$ and $1/n_{\Delta}$ are the relative variances of $g_2(\beta_1,\beta_2)$ along sum and difference variables $\beta_\Sigma$ and $\beta_\Delta$ respectively, and $\Delta (1/n)_{tot} \equiv 1/n_{\Sigma} - 1/n_{\Delta}$ is the relative {\em covariance} of $g_2$~\cite{2pttemp}, measuring velocity/temperature correlations. For the examples in panels (d), (e) and (f) of Fig.~\ref{Figure3},
$1/n_{\Sigma} = 1/n_{\Delta}$, $1/n_{\Sigma} > 0$ and $1/n_{\Delta}
= 0$, and $1/n_{\Sigma} > 1/n_{\Delta} > 0$, respectively. Mixed-event pair distribution $F_{mix}(p_{t1},p_{t2})$, a product of one-dimensional L\'evy distributions [Eq.~(\ref{Eq:XX1})], has the form of Eq.~(\ref{fsib}) but with $n_{\Sigma} = n_{\Delta} = n_{fluct}$.

Ratio
\bea
r_{model} \equiv F_{sib}/F_{mix},
\eea
referred to as a 2D L\'evy saddle,
predicts a saddle-shape when $g_2(\beta_1,\beta_2)$ has non-zero covariance and is the analytical quantity to be compared to data. It can be tested by comparison to the data in Fig.~\ref{Figure1} via chi-square fits.  We emphasize for this 2D case that any peaked function $g_2(\beta_1,\beta_2)$ with {\em nonzero covariance} results in a 2D saddle shape distribution for $r_{model}$.
The gamma distribution times gaussian 2D model for $g_2$ was chosen for mathematical convenience but it is reasonable given the form of the measured event-wise $\langle p_t \rangle$ distribution. The variance of $g_2$ along the difference direction $\beta_\Delta$ measures the average degree of equilibration of the events in the ensemble.
Relative variance {\em differences} $\Delta (1/n)_{\Sigma} \equiv (1/n_{\Sigma} - 1/n_{fluct})$ and $\Delta (1/n)_{\Delta} \equiv (1/n_{\Delta} - 1/n_{fluct})$ measure the saddle {\em curvatures} of $r_{model}$ (and hence the data) along sum and difference directions at the origin, and are the quantities best determined by these fits. Sensitivity to the magnitudes of the relative variances $1/n_{\Sigma}$ and $1/n_{\Delta}$ is discussed in the next section.

\section{Analytical Model Fits}
\label{Sec:Model}

\begin{figure}[t]
\includegraphics[keepaspectratio,width=1.65in]{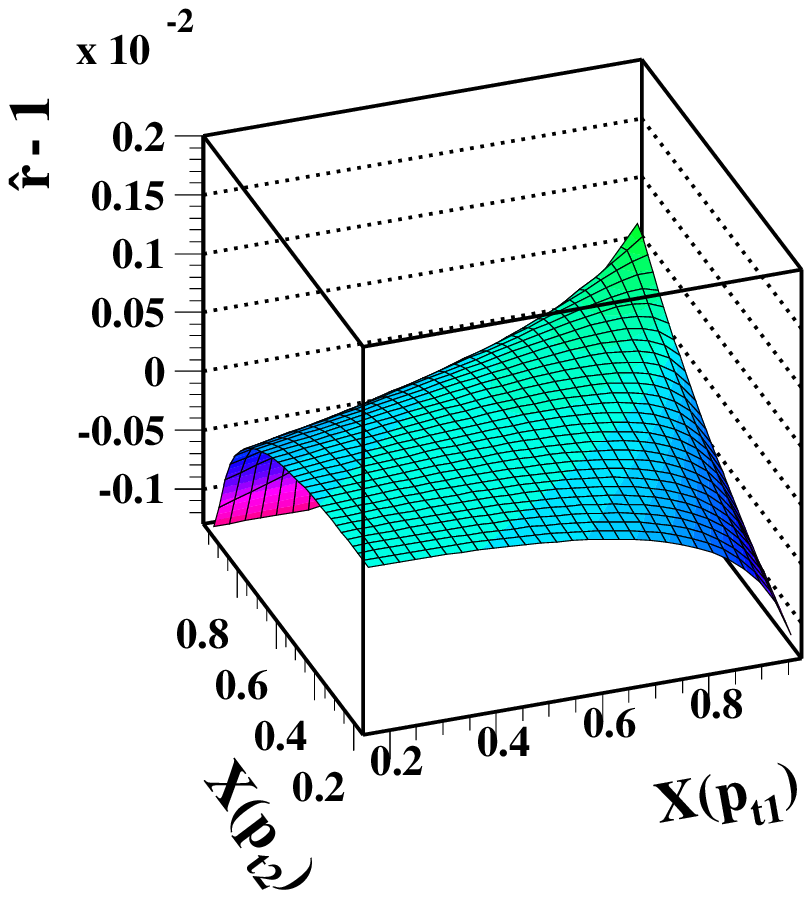}
\includegraphics[keepaspectratio,width=1.65in]{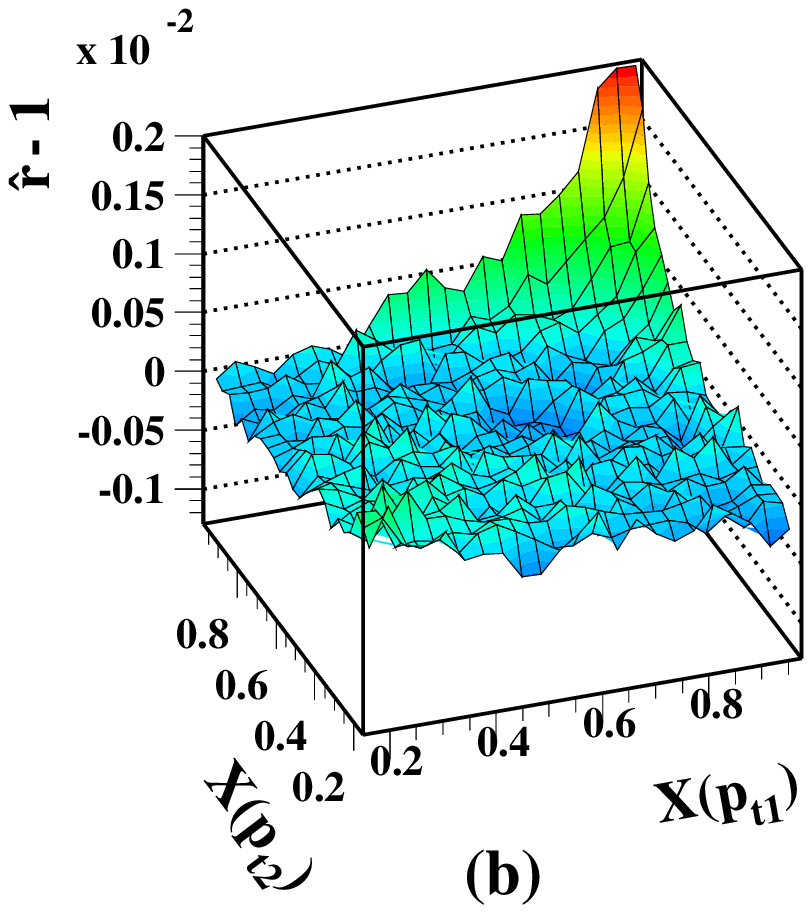}
\vspace{-.05in}
\caption{\label{Figure4} Left: pair-density net ratio $r_{model}[X(p_{t1}),X(p_{t2})] - 1$ for model fit to mid-central (b) Au-Au collisions. Right: residuals (data $-$ model) for mid-central collisions.}
\end{figure}

Data in Fig.~\ref{Figure1} (excluding peak region $X(p_t)_{\Sigma}
> 1.6$) were fitted with $r_{model} - 1 + {\tilde C}$ by varying parameters $n_{\Sigma}$, $n_{\Delta}$ and ${\tilde C}$ (offset). Parameters $\beta_0=5$~GeV$^{-1}$ and $m_0 = m_{\pi}$ were fixed by the (pion dominated) inclusive single-particle $p_t$ spectrum for $p_t <$ 1 GeV/$c$. The fits are insensitive to the absolute value of $1/n_{fluct}$; its value was fixed as follows.
Parameter $1/n$ when fitted to the single particle $m_t$ spectrum~\cite{spectra}, using an analog of Eq.~(\ref{Eq:XX1}) with $n_{fluct}$ replaced by $n$, accounts for the deviation between the measured distribution and e$^{-\beta m_t}$. 
In general, both collective
radial expansion velocity~\cite{ssh} and effective temperature
fluctuations contribute to the curvature (decreasing slope) of the $m_t$
spectrum relative to Boltzmann reference e$^{-\beta m_t}$ at increasing $m_t$ shown by
the dashed curve in Fig.~\ref{Figure3}(c). Both contributions are included in
parameter $n$ in Eq.(4), when fitted to the single particle distribution, resulting in an apparent variance,
$1/n$, given by an incoherent sum of contributions from radial flow,
$1/n_{flow}$, and effective temperature fluctations, $1/n_{fluct}$,
where $1/n = 1/n_{flow} + 1/n_{fluct}$. However, for the effective temperature fluctuation model developed in the preceding section only 
component $1/n_{fluct}$ is relevant to the 2D L\'evy saddle fit but it is not accessible because fits to correlation data ($\hat{r}-1$) poorly constrain absolute quantities $1/n_{\Sigma}$ and $1/n_{\Delta}$. However, {\em differences} $\Delta(1/n)_{\Sigma,\Delta}$ are well determined by the saddle curvatures, nearly independently of the assumed value of $1/n_{fluct}$ in $r_{model}$.
The maximum value for $1/n_{fluct}$ corresponds to $1/n = 1/13$ in the no-flow limit, $1/n_{flow} = 0$, where $n = 13$ is obtained from the L\'evy distribution fit to the single particle $m_t$ spectrum~\cite{spectra}. The minimum value of 0.0009 corresponds to that necessitated by the fitted values of $\Delta(1/n)_{\Delta}$ in Table~\ref{TableI} in the limit $1/n_{\Delta} \rightarrow 0$. The fits were insensitive to variations of $1/n_{fluct}$ in this range, intermediate value $1/n_{fluct} = 0.03$ near the center of the allowed range provided stable
$\Delta (1/n)_\Sigma$ and $\Delta (1/n)_\Delta$ fit values. Best-fit parameters and $\chi^2$/DoF for the saddle fits are listed in Table~\ref{TableI}.  The model function and residuals for the fit to centrality (b) are shown in Fig.~\ref{Figure4}.

\vspace{-.05in}

\begin{table}[h]
\caption{\label{TableI}
Parameters and fitting errors (only) for 2D velocity/temperature fluctuation 
model for each centrality bin, (a) - (d)
(central - peripheral
as in Fig.~\ref{Figure1}). Errors (last column) represent
fitting uncertainties. Systematic errors are 7-12\% \cite{norm}.
Mean multiplicities of used particles in the acceptance, $\bar N$, are listed for each centrality bin. 
Quantities ${\cal S}$ (last row) are correction factors  for contamination and tracking inefficiency~\cite{meanptprl}.}
\begin{tabular}{|c|c|c|c|c|c|} \hline
centrality  & (d) & (c)  & (b)  & (a)  &  error\footnote{Range of fitting errors
in percent from peripheral to central.}(\%)  \\
\hline \vspace{-.12in} & & & & & \\
$\bar{N}^{ }$  &   115.5   &   424.9   &  790.2   &   983.0 &  \\
\hline \vspace{-.12in} & & & & & \\
${\tilde C} \times 10^4$
                       & -11.6  & -0.820  &  0.787  &  0.750  &  6-14  \\
$\Delta(1/n)_{\Sigma} \times 10^4$
                       &  3.54  &  0.611  &  0.183  &  0.118  &  6-24  \\
$\Delta(1/n)_{\Delta} \times 10^4$
                       & -8.61  & -3.33   & -2.53   & -2.04   &  6-3   \\
$\Delta(1/n)_{tot} \times 10^4$
                       &  12.2  &  3.95   &  2.71   &  2.16   &        \\
$\chi^2$/DoF  &  $\frac{348}{286}$    &  $\frac{313}{286}$    &  $\frac{475}{286}$    &  $\frac{402}{286}$    &   \\
 \vspace{-.12in} & & & & & \\
\hline
\hline
${\cal S}$ &   1.19    &   1.22    &  1.25    &   1.27  &  8\footnote{Systematic
error.}  \\
\hline 
\end{tabular}
\end{table}

\begin{figure}[t]
\includegraphics[keepaspectratio,width=1.65in]{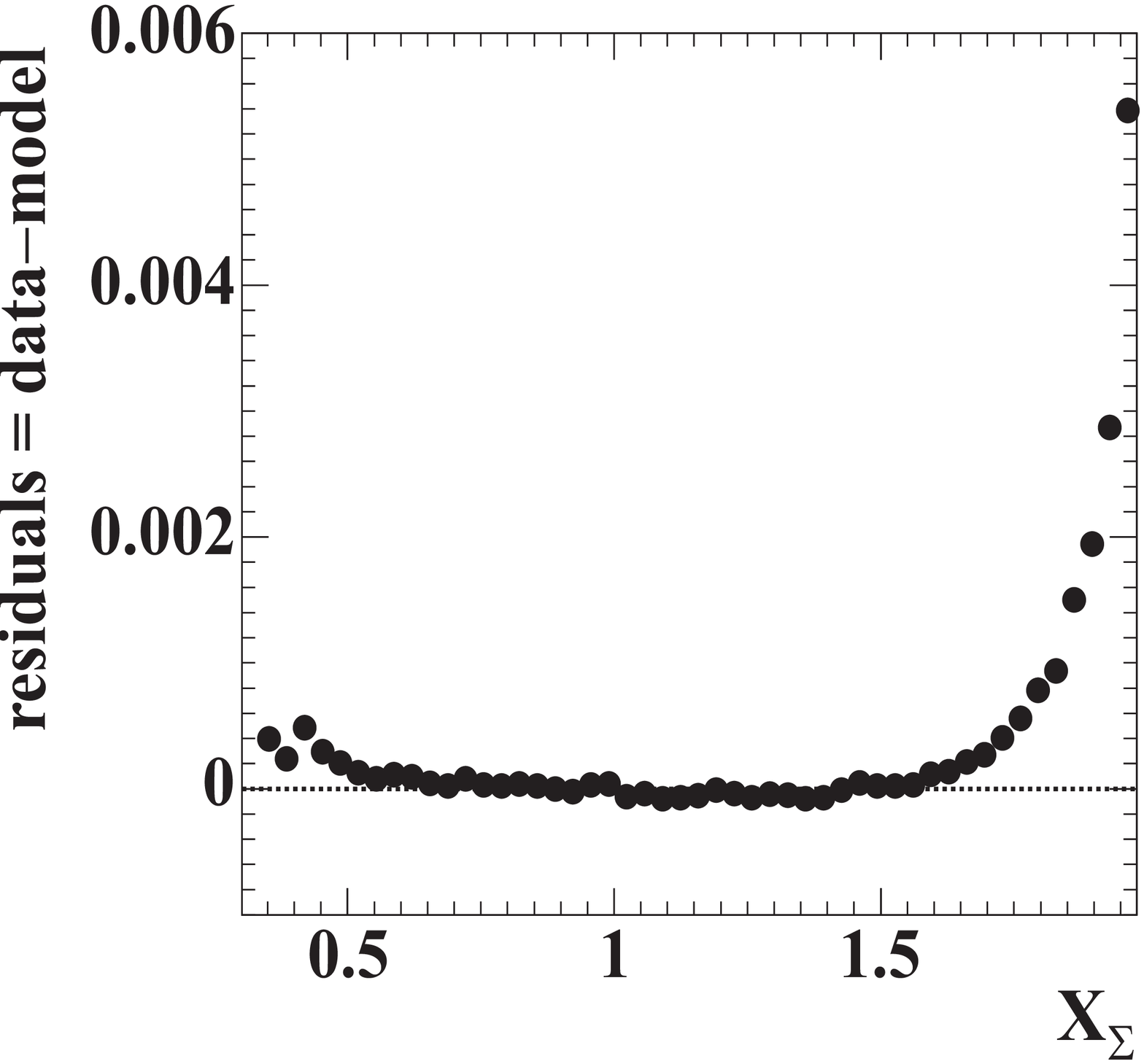}
\includegraphics[height=1.55in,width=1.65in]{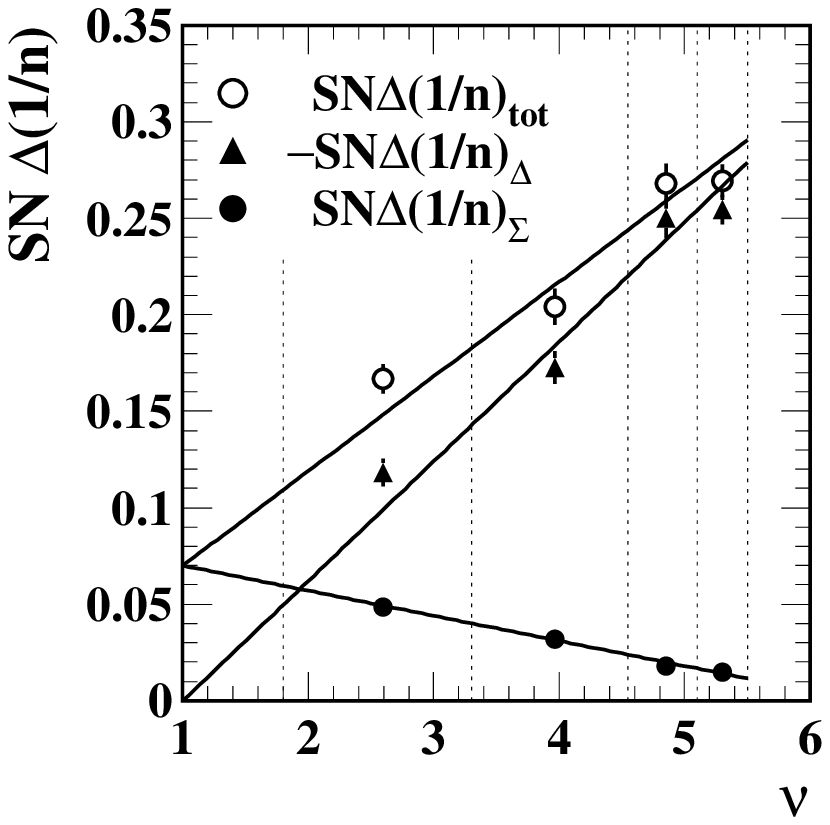}
\caption{\label{Figure5} 
Left: Residuals from 2D L\'evy saddle fit to mid-central (b) data in Fig.~\ref{Figure1} projected onto sum variable $X(p_t)_{\Sigma} = X(p_{t1}) + X(p_{t2})$.
Right: Efficiency-corrected {\em per-particle}\, saddle-curvature measures~\cite{eff} on centrality $\nu$: ${\cal S}\bar N\,  \Delta (1/n)_\Sigma$ (dots), 
$-{\cal S}\bar N \,  \Delta (1/n)_\Delta$
(triangles) and ${\cal S}\bar N\, \Delta (1/n)_{tot}$ (open circles).
Data symbols include fitting errors only~\cite{norm}. 
Solid lines are linear fits.
}
\end{figure}

Two-dimensional saddle-fit residuals, as in Fig.~\ref{Figure4} (right panel), are approximately constant along directions parallel to the
$X(p_t)_{\Delta} = X(p_{t1}) - X(p_{t2})$ axis for each value of
$X(p_t)_{\Sigma}$ and are small for $X(p_t)_{\Sigma} < 1.5$. The L\'evy temperature fluctuation model adequately describes the saddle structure. Residuals from the fit for mid-central events (b) are shown in Fig.~\ref{Figure5} (left panel) projected onto sum variable $X(p_t)_{\Sigma}$.  Errors are included in the data symbols and are smaller than those in Fig.~\ref{Figure4} (right panel) due to bin averaging. Residuals for other centralities are similar, but differ in amplitude.  We hypothesize that this residual structure is due to correlated final-state hadrons associated with initial-state semi-hard parton scattering~\cite{jeffp}.

Centrality dependences of efficiency-corrected model parameters ${\cal S}\bar N \Delta(1/n)$~\cite{eff}, which determine saddle-shape correlation amplitudes in Fig.~\ref{Figure1}, are shown in Fig.~\ref{Figure5} (right panel). The linear trends suggested by the solid lines are notable. Multiplication by factor ${\cal S}\bar N$ estimates correlation amplitudes per final state primary particle as discussed below. Centrality measure
$\nu$ estimates the mean participant path length as the average number of encountered nucleons per participant nucleon in the incident nucleus. For this analysis $\nu \equiv 5.5 \, (N/N_0)^{1/3} \simeq 5.5\, (N_{part}/N_{part,max})^{1/3} \simeq 2N_{bin}/N_{part}$, based on Glauber-model simulations where
$N_{part}$ ($N_{bin}$) is the number of participant nucleons (binary collisions). 

The reasons for multiplying the parameters in Table~\ref{TableI} by ${\cal S}$ and $\bar N$ are the following.  Multiplication of $(\hat r - 1)$ by $\bar N$ yields the density of correlated pairs {\em per final-state particle}~\cite{axialcd}, typically $O(1)$ for all centralities.  $\bar N(\hat r - 1)$ would be independent of centrality if Au-Au collisions were linear superpositions of p-p collisions (participant scaling) because the amplitude of the numerator of $(\hat r - 1)$, which is proportional to the density of correlated pairs, would scale with participant number, or in this model with $\bar N$, while the denominator is proportional to $\bar N^2$. Therefore variation of $\bar N(\hat r - 1)$ with centrality directly displays the effects of those aspects of Au-Au collisions which do not follow na\"ive p-p superposition.
Factor ${\cal S}$ is defined as the ratio of true, primary particle yield ({\em i.e.,} 100\% tracking efficiency and no background contamination) estimated for these data in Ref.~\cite{spectra} divided by the actual multiplicity used in this analysis corrected for the $\sim 7$\% background contamination.  ${\cal S}$ is essentially the reciprocal of the charged-particle tracking efficiency, specific for the present analysis.  Multiplication by factor ${\cal S} \bar N$ of the parameters in Table~\ref{TableI} therefore estimates the correlation amplitudes per final-state particle for 100\% tracking efficiency and no background contamination, assuming the measured correlations include background-primary particle correlations half-way between the limits described in Sec.~\ref{Sec:Errors}. The uncertainty in extrapolating to the true primary particle yield is estimated to be 8\%, most of which is due to the 7\% systematic uncertainty in the measured charged hadron yield~\cite{spectra}.  The combined systematic uncertainty for the efficiency corrected amplitudes is from $\pm 11-14$\% across the $X(p_{t1})$ {\em vs} $X(p_{t2})$ space.

\section{Discussion}
\label{Sec:Diss}



Correlations on $p_t$ have two main components, a saddle shape and a peak at higher $p_t$. By measuring the saddle curvatures we infer the relative covariance of two-point distribution $g_2(\beta_1,\beta_2)$ and hence the average two-point correlation amplitude of the temperature/velocity structure of the composite particle source. We now consider possible dynamical origins of that structure.

The analysis of the saddle-shape produces accurate results for relative variance differences $\Delta(1/n)_{\Sigma} = (\sigma^2_{\beta_\Sigma} - \sigma^2_{\beta})/\beta^2_0$, $\Delta(1/n)_{\Delta} = (\sigma^2_{\beta_\Delta} - \sigma^2_{\beta})/\beta^2_0$, and the corresponding $\Delta(1/n)_{tot} = (\sigma^2_{\beta_\Sigma} - \sigma^2_{\beta_\Delta})/\beta^2_0$ for effective temperature fluctuations. The measurements do not constrain the absolute magnitudes of the individual variances, $\sigma^2_{\beta_\Sigma}$ and $\sigma^2_{\beta_\Delta}$. The minimum possible values, consistent with the saddle-shape conditions and the single-particle $m_t$ spectra, correspond to $\sigma^2_{\beta_\Delta} = 0$ and $1/n_{fluct} = -\Delta(1/n)_{\Delta}$, resulting in $\sigma_\beta / \beta_0 \cong \sigma_T/T_0 = 1.4$\% to 2.9\% global event-to-event temperature/velocity fluctuation from central to peripheral collisions, respectively. In this case $1/n \cong 1/n_{flow}$ and global temperature/velocity fluctuations contribute negligibly to the upward curvature of the $dN/m_tdm_t$ spectrum.  The maximum values for the variances correspond to $1/n_{flow} = 0$, resulting in $1/n_{fluct} = 1/n$ and $\sigma_\beta / \beta_0 \cong \sigma_T/T_0 = \sqrt{1/n} = 30$\%, where $\sigma_{\beta_\Delta} \sim \sigma_{\beta}$, corresponding to 30\% local temperature/velocity fluctuations within each event, a significantly non-equilibrated system. Thus, local temperature variation could range between 0 and 30\%. One can ask what is the source of the fluctuating effective temperature, and is local source velocity rather than temperature a more appropriate quantity?

Given the correlation peaks at higher $p_t$ it is reasonable to offer the hypothesis that the saddle-shape correlation structure in Fig.~\ref{Figure1} results from in-medium modification, specifically momentum dissipation on $(p_{t1},p_{t2})$ of a two-particle distribution from fragmenting, semi-hard scattered partons in the initial-stage of the collision. Since no selection was made on leading particle or high-$p_t$ ``trigger'' particle for these data we refer to the hadrons associated with a semi-hard, initial-state scattered parton as a {\em minijet}~\cite{jetquench,minijet}. Minijet production in Au-Au collisions should increase approximately linearly with $N_{bin}$~\cite{raa,raaother} while the subsequent momentum dissipation should monotonically increase with greater minijet production. Correlation amplitudes per final state particle (the latter approximately proportional to $N_{part}$) should therefore increase monotonically with mean participant path length $\nu \cong N_{bin}/(N_{part}/2)$, thus providing a basis for experimental tests of this hypothesis.

The linear trends in  Fig.~\ref{Figure5} (right panel) therefore support, but do not {\em require}, a minijet $-$ momentum dissipation mechanism for the observed correlations on $p_t$.
%
%
In Fig.~\ref{Figure5} we also observe 1) reduced curvature along the sum direction and 2) increased curvature along the difference direction which may represent respectively transport of semi-hard parton structure to lower $p_t$ and a more correlated bulk medium. Given a minijet interpretation of ${\cal S}\bar{N}\,\Delta(1/n)_{tot}$, the combined trends 1) and 2) represent strong evidence for increased parton dissipation in the more central Au-Au collisions. The present results complement the observed suppression of high-$p_t$ spectra ($R_{AA}$)~\cite{raa,raaother} and suppression of large angle trigger-particle -- associated-particle conditional distributions on $\Delta\phi$~\cite{backjet,phenixbackjet} in central Au-Au collisions at RHIC. It is very likely that the lower-$p_t$ fluctuations and correlations reported here are, at least in large part, a consequence of the processes which lead to the above suppressions at higher-$p_t$.

It is important to note that these correlations on transverse momentum observed at relatively low $p_t$ reveal nominally `soft' structure in relativistic heavy ion collisions which scales with the number of binary collisions $N_{bin}$, whereas a low-$p_t$ inclusive quantity such as multiplicity scales with participant number $N_{part}$. Binary-collision scaling is conventionally thought to be an aspect of high-$p_t$ physics and initial-state scattering. This analysis suggests that substantial effects of initial-state parton scattering are manifest at low $p_t$ in more central heavy ion collisions.




\section{Summary}
\label{Sec:Sum}

In conclusion, the dynamical origins of excess $\langle p_t \rangle$ fluctuations
in Au-Au collisions at RHIC are studied in this analysis of two-particle correlations on $(p_{t1},p_{t2})$. The velocity/temperature structure of heavy ion collisions suggested by these correlations is unanticipated by theoretical models~\cite{jetquench,rqmd,ampt}.  Lacking in these models is the simultaneous inclusion of hard scattering in the initial state with subsequent medium modification of the fragmentation function and/or interactions between the medium and the hadrons associated with the scattered partons. Nevertheless it seems plausible to interpret the observed correlations on $(p_{t1},p_{t2})$ as resulting from this sort of semi-hard parton scattering and subsequent medium modified fragmentation and/or associated hadron distributions on $p_t$ in the more central Au-Au collisions. In this picture, with increasing centrality the transverse momentum associated with the two-particle fragment distribution from initial-state semi-hard parton scattering is shifted to lower $p_t$, asymptotically approaching a form consistent with random velocity/temperature variations (L\'evy saddle) as a manifestation of substantial but incomplete equilibration. These newly-observed $p_t$ correlations may thus reveal minijet dissipation in the medium produced by Au-Au collisions at RHIC. 


We thank the RHIC Operations Group and RCF at BNL, and the
NERSC Center at LBNL for their support. This work was supported
in part by the Offices of NP and HEP within the U.S. DOE Office 
of Science; the U.S. NSF; the BMBF of Germany; IN2P3, RA, RPL, and
EMN of France; EPSRC of the United Kingdom; FAPESP of Brazil;
the Russian Ministry of Science and Technology; the Ministry of
Education and the NNSFC of China; IRP and GA of the Czech Republic,
FOM of the Netherlands, DAE, DST, and CSIR of the Government
of India; Swiss NSF; the Polish State Committee for Scientific 
Research; SRDA of Slovakia, and the Korea Sci. \& Eng. Foundation.


\begin{thebibliography}{99}
\bibitem{Xmap}
Function $X(p_t)$ permits a fully analytic
description of the data in $[X(p_{t1}),X(p_{t2})]$ space using the fitting
model in $(p_{t1},p_{t2})$ space described in Sec.~\ref{Sec:Dist}. 
Choice $m_0 = m_{\pi} \approx T$ 
emphasizes the soft part of the $m_t$ spectrum in mapping from $p_t$ to $X(p_t)$ where a temperature model is more appropriate.
\bibitem{rhomix}
The weighted integral of $\rho_{mix}$ corresponds to
$\bar N$ times the second term
in the first line of Eq.~(\ref{Eq2}). The event
ensemble average number of mixed-event pairs for that term is
$\overline{N(N-1)} =
\bar{N}^2 - \bar{N} + \sigma_N^2$, where $\sigma_N^2$ is the
variance of the multiplicity distribution, assumed to be Poisson,
such that $\sigma_N^2 = \bar{N}$.
\bibitem{trackcuts}
Two-track merging cuts required average separation distance $\geq 10$~cm based on two-track separations at nine radial positions in the TPC.  Two tracks which cross within the TPC with separations less than 10~cm ($z$) and 30~cm (azimuth) at mid-radius from the TPC axis were also excluded.
\bibitem{endpoint}
$N_0$, the half-maximum point at the end of the minimum-bias distribution plotted as $d\sigma / dN^{1/4}$, is an estimator on $N$ for the maximum number of participants; $N_{part}/N_{part,max} \simeq N/N_0$ within 4\%.
\bibitem{tempflow} Analysis of the measurements presented here cannot distinguish between fluctuating temperature or flow velocity which would require identified particle mass at higher $p_t$ and over the large $\eta,\phi$ acceptance of the STAR TPC.
\bibitem{norm}
Systematic errors for quantities in Fig.~\ref{Figure5} (right panel), due to systematic uncertainties in the data (7-12\%) plus background and efficiency corrections (8\%), are 11-14\%.
\end{thebibliography}
\end{document}